%% file: paper.tex
\documentclass[useAMS,usenatbib]{mn2e}
\usepackage{aas_macros}
\usepackage[a4paper,centering, totalwidth=520pt, totalheight=700pt]{geometry}
\input{local-commands}
\usepackage[fleqn]{amsmath} 
\usepackage{graphicx}
\usepackage{amssymb}
\usepackage{amsmath}
\usepackage{enumerate}
\usepackage{microtype}
\usepackage{color}

\title[Cosmic Velocity Fields]{The Properties of Cosmic Velocity Fields}
\author[O. Hahn, R. Angulo \& T. Abel ]{Oliver Hahn$^{1}$\thanks{Email:
    hahn@phys.ethz.ch}, Raul E. Angulo$^{2}$ \& Tom Abel$^{3}$ \\
  $^{1}$Department of Physics, ETH Zurich, CH-8093 Z\"urich,
  Switzerland \\
  $^{2}$Centro de Estudios de Fisica del Cosmos de Arag\'on, 
  San Juan, 1 Planta-2, 44001 Teruel, Spain\\
  $^{3}$Kavli Institute for Particle Astrophysics and Cosmology, \\ 
  \quad Stanford University, SLAC National Accelerator Laboratory,
  Menlo Park, CA 94025, USA\\
    }

\begin{document}

\date{submitted to MNRAS}
\pagerange{\pageref{firstpage}--\pageref{lastpage}} \pubyear{2014}
\maketitle

\label{firstpage}

\begin{abstract}
Understanding the velocity field is very important for modern cosmology: it gives insights to structure formation in general, and also its properties are crucial ingredients in modelling redshift-space distortions and in interpreting measurements of the kinetic Sunyaev-Zeldovich effect. Unfortunately, characterising the velocity field in cosmological $N$-body simulations is inherently complicated by two facts: i) The velocity field becomes manifestly multi-valued after shell-crossing and has discontinuities at caustics. This is due to the collisionless nature of dark matter. ii) $N$-body simulations sample the velocity field only at a set of discrete locations, with poor resolution in low-density regions. In this paper, we discuss how the associated problems can be circumvented by using a phase-space interpolation technique. This method provides extremely accurate estimates of the cosmic velocity fields and its derivatives, which can be properly defined without the need of the arbitrary ``coarse-graining'' procedure commonly used. We explore in detail the configuration-space properties of the cosmic velocity field on very large scales and in the highly nonlinear regime. In particular, we characterise the divergence and curl of the velocity field, present their one-point statistics, analyse the Fourier-space properties and provide fitting formulae for the velocity divergence bias relative to the non-linear matter power spectrum. We furthermore contrast some of the interesting differences in the velocity fields of warm and cold dark matter models. We anticipate that the high-precision measurements carried out here will help to understand in detail the dynamics of dark matter and the structures it forms.
\end{abstract}


\maketitle

\section{Introduction}
\noindent $N$-body simulations in computational cosmology
\citep[e.g.][]{Melott:1982, Efstathiou:1985, Peebles:1989,Springel:2005,
Angulo:2012} are an invaluable tool to study the dynamics of cosmic dark matter
over time. Statistics of the dark matter velocity field are important for
cosmological measurements of redshift-space distortions from galaxy surveys
\citep[see e.g.][for a classic review]{Dekel:1994}, of the kinetic
Sunyaev-Zeldovich effect, as well as for more theoretically motivated questions
regarding the formation and dynamics of the cosmic large-scale structure
\citep[e.g.][]{Bertschinger:1989, Bernardeau:1996, Pichon:1999,Pueblas:2009,
Kitaura:2012}. 

Unfortunately, there is a fundamental problem when determining the dark
matter velocity field from $N$-body simulations: while the velocity field is
defined everywhere in space in the continuous limit, in simulations it is sampled only at the
(mass-weighted) particle positions. A recent discussion by \cite{Jennings:2011}
highlights the potentially large uncertainties on different measures of the
velocity power spectrum from cosmological $N$-body simulations. Using a standard
cloud-in-cell  \citep[CIC][]{Hockney:1981} deposit to project simulation
results on a grid leads to unacceptable levels of noise and biases, severely complicating
a reliable measurement of this quantity.

Various approaches can be found in the literature to reconstruct a continuous velocity
field and address this sampling issue. One possibility is to apply a large-scale
smoothing (i.e. using a kernel estimator) with a given fixed or adaptive kernel function
\citep[e.g.][]{Bertschinger:1989, Melott:1993}. This has advantages when
applied to galaxy catalogs and the interest is mostly limited to the large scales that
are still close to the linear regime and where shell-crossing can be neglected.
Another approach is based on using inverse SPH-smoothed
velocity fields \citep{Colombi:2007}. A third class of approaches employs 
tessellations of the spatial distribution of sampling points which yields a good estimator
in the absence of shell-crossing. Tessellations allow to define unique volumes around
particles (by Voronoi tessellations) or by connecting the particles \citep[Delaunay
triangulations: e.g.][]{Icke:1987, Bernardeau:1996,Pueblas:2009}, and they
have been particularly popular in recent analyses of $N$-body simulations
\citep[e.g.][]{Pandey:2012}. As shown by \cite{Bernardeau:1996} and
\cite{Pueblas:2009}, Delaunay tessellations of the particle distribution can be
used to better control noise and measurement errors of velocity power spectra
from $N$-body simulations. In addition, \cite{Jennings:2012} showed that the DTFE
estimator \citep{Schaap:2000,Pelupessy:2003, Schaap:2007}, as implemented by
\cite{Cautun:2011}, gives much more reliable spectral properties compared to
fixed kernel smoothing (confirming a similar finding by
\citealt{Pueblas:2009}). 

Despite recent progress, while trying to solve the two issues described above, these
methods sacrificed a direct measurement of the velocity field by that of a
smoothed field and/or introduced high levels of noise. This is particularly
important: due to the collisionless nature of dark matter, gravitational
collapse leads to a multi-valued velocity field in multi-stream regions (quite
in contrast to the behaviour of an ideal fluid), and discontinuities appear in the
velocity field. Therefore, as we will discuss throughout this paper, the properties 
of the volume-averaged (i.e. ''coarse-grained'') velocity field are not identical 
to the properties of the mean velocity field.
For instance, a non-zero vorticity appears in the coarse-grained velocity field
even in regions with particles whose orbits have not 
yet crossed. Additionally, the properties of the coarse-grained field strongly 
depend on the (arbitrary) scale on which the volume average is performed.

In this paper, we will demonstrate how the mean 
velocity field and its differential properties (specifically, its vorticity and divergence) 
can be computed avoiding the problems outlined above. In particular, we will propose
a scheme that provides an accurate (and to our knowledge the only) way to determine the
mean velocity field without coarse-graining and in the presence of multi-streaming. 
This is a direct consequence of the possibility to define an explicit projection from phase-space  
into configuration space based on a reconstruction of the fine-grained distribution function. Therefore, and
in contrast to coarse graining, there is no arbitrary scale in this direct projection. 

Our approach is based on the method to analyse cosmological $N$-body simulations
proposed by \cite{Abel:2012} (AHK12 hereafter). The key idea rests on the fact
that structure formation in cold dark matter starts out on a very thin sheet in
configuration space with an almost infinitesimally small extent in velocity
space. Consequently, one can think of the modelled particles as the moving
(massless) vertices of a tessellation of this initial (Lagrangian) phase space sheet.
\cite{Shandarin:2011} independently developed the same idea with only minor
differences in the implementation. The estimates of densities, velocities and
velocity dispersion therefore incorporate the information of neighbouring
particles in phase-space. Such a Lagrangian
tessellation is able to accurately represent anisotropic deformations of the
density field which allows it to be free from the artificial clumping seen in
(adaptive) kernel smoothing~\citep{Hahn:2012, Angulo:2013}.

 The features described above make this method 
unique from all other methods such as adaptive kernel smoothing (as e.g. in SPH
\citealt{Monaghan:1992}), CiC deposits or DTFE. We also note that our method is 
distinct from the phase space estimation techniques
of \cite{2006MNRAS.373.1293S} or \cite{2010CoPhC.181.1438A}, which
treat $N$-body data results as Monte Carlo sampled realisations of the
micro-physical phase space structure of the dark matter distribution.
Consequently all these methods are subject to noise even in the case of the
heavily smoothed SPH method \citep[see the detailed discussion
in][]{Hahn:2012}. 

Using our explicit phase-space projection procedure applied to
dark matter $N$-body simulations,
we are able to (i) show how the velocity
field switches from convergent to divergent flow in multi-stream regions, with
a remnant convergent core in the centres of high-density structures such as haloes
and filaments; (ii) demonstrate that vorticity is a multi-stream phenomenon, which 
peaks at caustics; (iii) present the 1-point statistics of the velocity divergence 
and vorticity, and (iv) provide high-resolution Fourier-space properties of the 
velocity field, in terms of velocity divergence and vorticity power spectra, and 
density-divergence cross-spectra. We also provide fitting formulae for the velocity divergence
bias relative to the non-linear matter power spectrum. We furthermore contrast some of
the interesting differences in the velocity fields of warm and cold dark matter
models. All our results are presented in Sections~\ref{sec:realspaceresults}
and \ref{sec:kspaceresults}.

The structure of this paper is as follows: We first discuss how mean velocity
fields are defined through a projection operation and give special attention to
derivatives of the projected velocity field that are non-trivial to compute due
to the discontinuous nature of the field in Section~\ref{sec:method}. Next, in
Section~\ref{sec:sims}, we introduce the $N$-body simulations used in this
work. In Section~\ref{sec:realspaceresults}, we present an analysis of the
real-space properties of the velocity field. In
Section~\ref{sec:kspaceresults}, we focus on the spectral properties of
velocity fields computed with our method and provide fits for the Fourier space
$k$-dependent bias of the velocity divergence relative to the cosmic density
field. We summarise our results in Section~\ref{sec:summary}.


\section{Projections onto configuration space, derivative operators and differentials of vector fields}
\label{sec:method}
In this section, we discuss how to obtain mean velocity fields from $N$-body
simulations using the cold dark matter sheet by projecting the fine-grained
distribution function from phase space onto configuration space. We then
discuss the differential properties of such projected fields. We will argue
that these differentials cannot in general be approximated by finite
differences and derive the correct expressions for the divergence and the curl
of the projected velocity field. Finally, we discuss how a piecewise linear
approximation to the fine-grained distribution function based on $N$-body
particles, as introduced in AHK12, can be used to compute the velocity field
and its differentials from simulations.


\subsection{The Vlasov-Poisson system and the distribution function of cold
fluids} We are concerned here with the mean velocity field of cold dark matter,
a cold collisionless self-gravitating fluid. The evolution of such a fluid is
fully described by the phase-space distribution function
$f(\mathbf{x},\mathbf{v},t)$ governed by the Vlasov-Poisson system of equations
\citep[see e.g.][the latter for a historical discussion]{Peebles:1980,Henon:1982} 

\begin{eqnarray}
0=\frac{{\rm d} f(\mathbf{x},\mathbf{v},t)}{{\rm d}t} &=& \frac{\partial f}{\partial t} + \frac{\mathbf{v}}{a^2}\cdot\boldsymbol{\nabla}_{x}f - \boldsymbol{\nabla}_{x}\phi\cdot\boldsymbol{\nabla}_{v} f, 
\label{eq:boltzmann} \\
\boldsymbol{\nabla}_{x}^{2} \phi & = & \frac{4\pi G}{a^3} \int {\rm d}^3v \, \left( f -  \bar{\rho} \right), 
\label{eq:Poisson}
\end{eqnarray}
where $G$ is the gravitational constant, $a$ is the cosmological scale
factor that itself obeys the first Friedmann equation, and $\bar{\rho}$ is the
mean density. The full phase space
distribution function $f(\mathbf{x},\mathbf{v},t)$ is in general manifestly
six-dimensional. For a perfectly cold fluid however, the distribution function
occupies only a three-dimensional hypersurface (in fact a sub-manifold) of
six-dimensional phase space \citep[see e.g. the discussion
in][]{Abel:2012,Shandarin:2011}. We can thus express the distribution function
through a parametrization of the hypersurface
$\mathbb{R}^3\to\mathbb{R}^6:\,\mathbf{q}\mapsto\left(\mathbf{x_q},\mathbf{v_q}\right)$
in terms of ``Lagrangian coordinates'' $\mathbf{q}$. The particles of an
$N$-body simulation can be thought of as a finite sampling
$\mathbf{q}_i\in\mathbb{R}^3,\,i=1\dots N$ of this hypersurface.

We note that the density is given by a projection of the distribution function
itself onto configuration space $\rho(\mathbf{x},t) \equiv \int {\rm d}^3v
f(\mathbf{x},\mathbf{v},t)$. In the case of a cold fluid, the integral over
velocity space becomes a discrete sum over the streams overlapping a given
point. For the notation, we omit the explicit time dependence in what follows.


\subsection{The mean velocity field}
\label{sec:coarse_grained_vel}
\begin{figure}
\begin{center}
\includegraphics[width=0.45\textwidth]{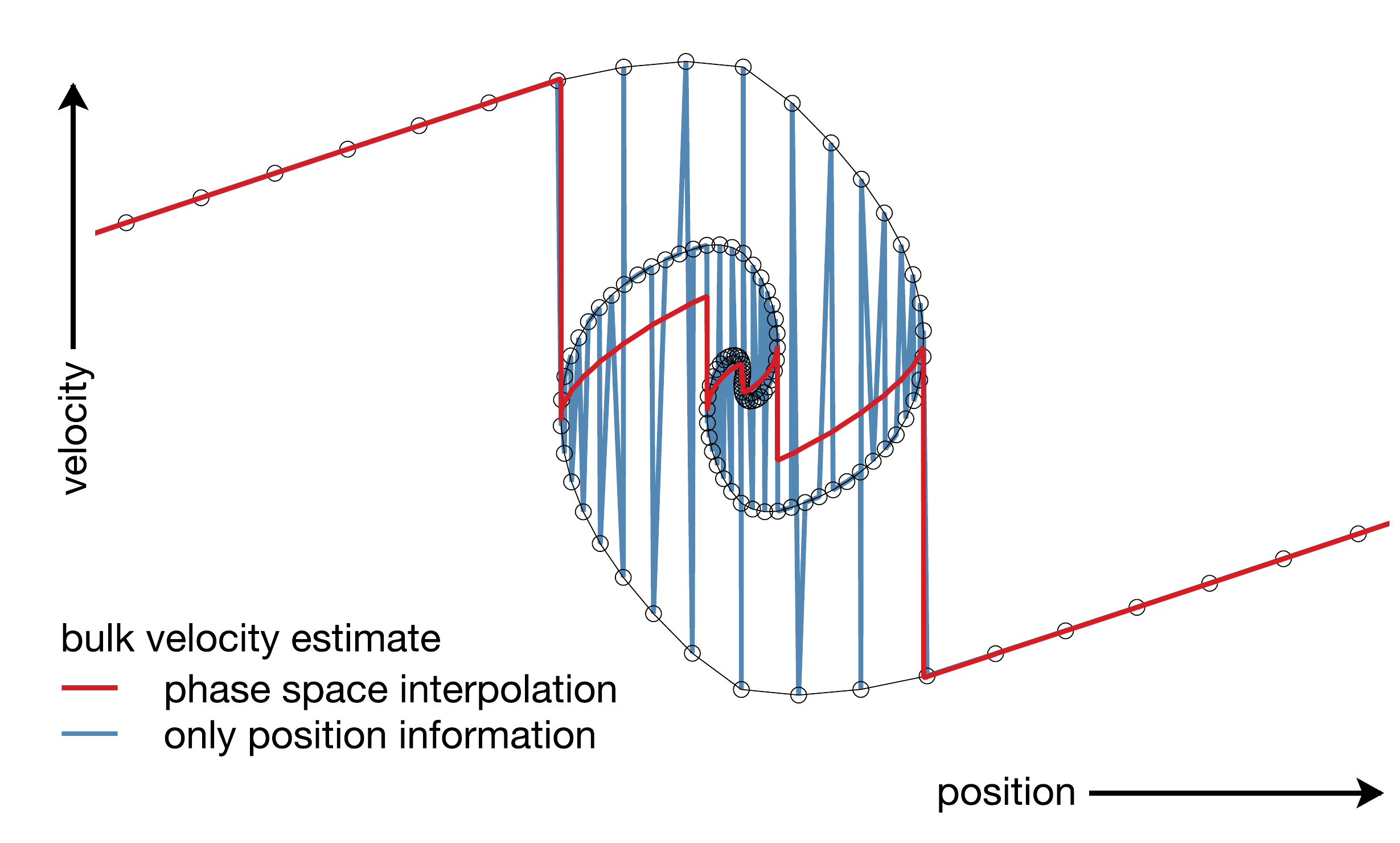}
\end{center}
\caption{\label{fig:velocity_estimators}Using tessellations to measure cosmic
velocity fields: Estimation of the mean velocity in a one-dimensional
plane wave collapse problem when only position information is available for the
tessellation (blue), or when the full phase space information can be used
(red). Black circles show the positions of the particles used in the
simulation, black lines indicate their connectivity on the dark matter sheet
and thus provide a piecewise linear approximation to the fine-grained
distribution function.  }
\end{figure}

The mean velocity is obtained from the first velocity moment of the
distribution function, i.e. the following projection onto configuration space
\begin{equation}
\left< \mathbf{v} \right> \equiv \frac{\int_{\mathbb{R}^3} \mathbf{v}\, f(\mathbf{x},\mathbf{v})\, {\rm d}^3 v}{\int_{\mathbb{R}^3} f(\mathbf{x},\mathbf{v}) \,{\rm d}^3 v} = \frac{\sum_{s\in{\rm S}(\mathbf{x})} \mathbf{v}_s(\mathbf{x})\,\rho_s(\mathbf{x}) }{\sum_{s\in{\rm S}(\mathbf{x})} \rho_s(\mathbf{x})},
\label{eq:vel_average}
\end{equation}
where ${\rm S}(\mathbf{x})$ is the set of all streams $s$ that contain point $\mathbf{x}$, a
subscript $s$ indicates the value of a field at $\mathbf{x}$ on a given sheet, and
the second equality holds for cold fluids, as discussed above. The mean
velocity field is thus given by a density-weighted average over the
multi-stream velocity field (cf. Fig.~\ref{fig:velocity_estimators}). We want
to remark explicitly that our projection operator $\left<\cdot\right>$ does
{\em not} involve the convolution with a smoothing kernel as is often done.
We note also that this projection operator, unlike kernel smoothing, is explicitly
idempotent, i.e. $\left< \left< \cdot \right> \right> = \left< \cdot \right>$.

\subsection{Differentials of velocity fields}
\label{sec:derivatives_general}
The differential flow properties of fluids are typically discussed in terms of the various components that
contribute to the first derivative -- i.e. the velocity gradient tensor $\boldsymbol{\nabla}\otimes\mathbf{v}$ -- 
of the the flow velocity field. This velocity gradient tensor is then further decomposed into (1) its trace, the velocity
divergence ${\rm div}\,\mathbf{v} = \boldsymbol{\nabla}\cdot\mathbf{v}$, (2) the anti-symmetric part of the tensor yields 
the vorticity vector $\boldsymbol{\omega}={\rm curl}\,\mathbf{v} = \boldsymbol{\nabla}\times\mathbf{v}$, and (3) the symmetric 
trace-free part of the tensor is the velocity shear $S_{ij} = \frac{1}{2}\left(\boldsymbol{\nabla}\otimes\mathbf{v} + (\boldsymbol{\nabla}\otimes\mathbf{v})^T\right)$, which we will however not consider further in this paper. The scalar magnitude of the vorticity vector field we call
simply ``vorticity'' $\omega = \left\| \boldsymbol{\omega} \right\|$.

We note that vorticity does not appear before shell-crossing if the
velocity field is not sourced by a vector potential and vorticity is initially
zero (this can also be seen simply as a consequence of Helmholtz's circulation theorem).
If initial vorticity is present, it is conserved and will be amplified by gravitational
collapse \citep[cf.][]{Buchert:1992}, but typically vortical initial modes are ignored, as we will 
also do in this paper.
Then, in the case of Newtonian gravity, the acceleration is given by the
gradient of the gravitational potential $\phi$. Hence, the single stream motion
of a Lagrangian fluid element $(\mathbf{x_q}, \mathbf{v_q})$ is curl-free
non-perturbatively at all times since at fixed $\mathbf{q}$: 
$\frac{{\rm d}}{{\rm d}t}\boldsymbol{\nabla} \times \mathbf{v_q} = -\boldsymbol{\nabla} \times
\left.\boldsymbol{\nabla}\phi\right|_{\mathbf{x_q}} =0$ if it is irrotational
initially \citep[cf. also][]{Bernardeau:2002}. After shell-crossing, various
fluid elements overlap and vorticity emerges \citep[cf. also][]{Pichon:1999},
as we will discuss further below.


\subsection{Differentials of the projected multi-stream velocity field}
\label{sec:derivatives}
\begin{figure}
\begin{center}
\includegraphics[width=0.2\textwidth]{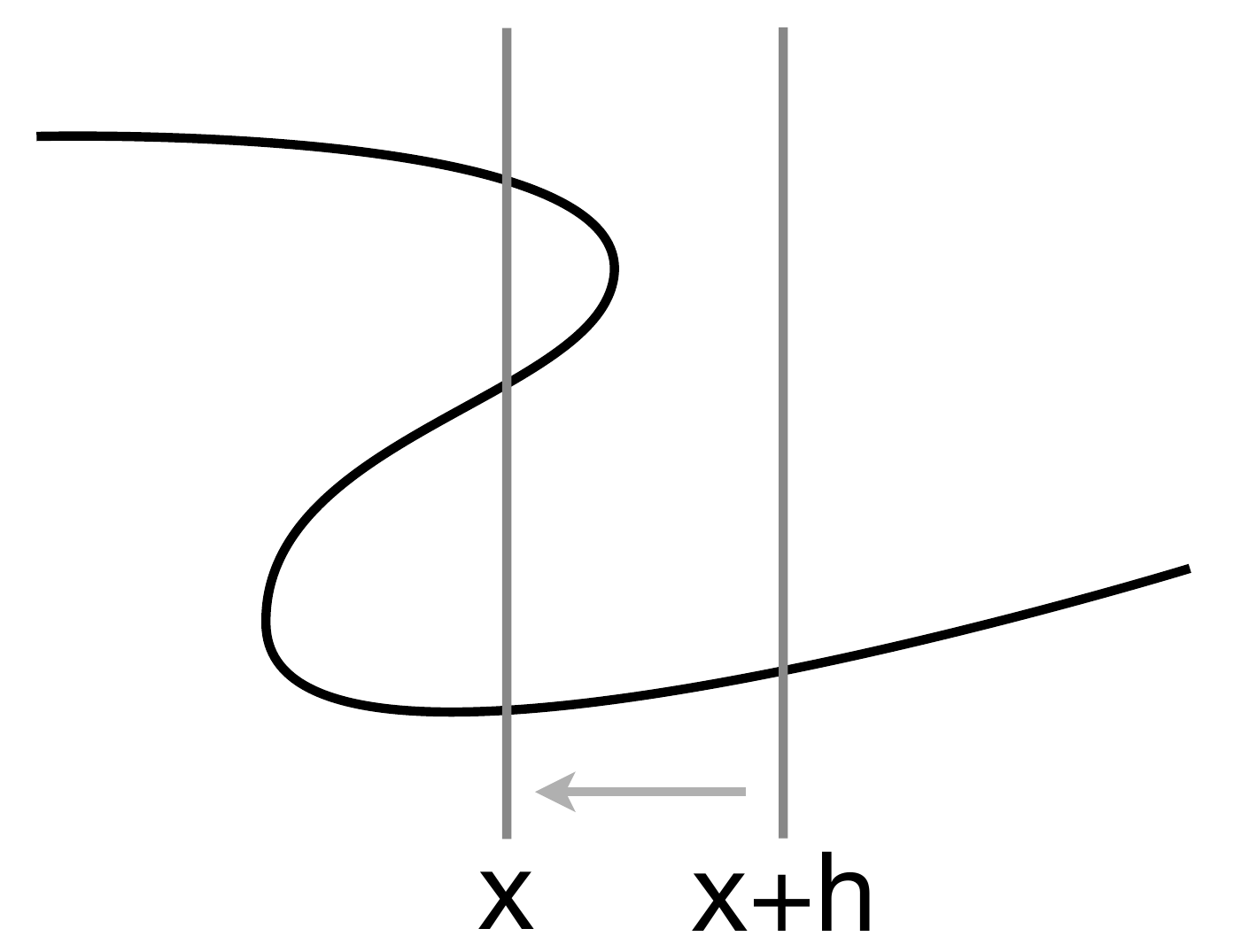}
\end{center}
\caption{\label{fig:caustic_limit}Illustration of a small piece of a cold
distribution function in 1+1 dimensional phase space (black line). The vertical
direction represents velocity, while the horizontal direction represents the
spatial dimension. When computing spatial derivatives of projected fields, the
difference evaluated on a finite interval does not approach the derivative if
the interval contains a caustic which implies a change in the number of streams
and a discontinuity in the respective mean velocity field.}
\end{figure}

As we will see next, the differentials of a projected variable are
singular at the location of caustics where the number of streams changes and
the projected field has a discontinuity. 
This can be easily seen
directly from the definition of a derivative applied to a projected field $\left<g\right>$:
\begin{equation}
\frac{{\rm d}\left< g(x) \right>}{{\rm d}x}  = \lim_{h\to 0} \frac{\left<g(x+h)\right>-\left<g(x)\right>}{h}.
\end{equation}
The right-hand-side can be written as
\begin{equation}
\lim_{h\to 0} \frac{1}{h} \frac{\sum_{s\in{\rm S}_1}\sum_{t\in{\rm S}_2}\rho_s(x+h)\rho_t(x)\left(g_s(x+h)-g_t(x)\right)}{\sum_{s\in{\rm S}_1}\sum_{t\in{\rm S}_2}\rho_s(x+h)\rho_t(x)},
\end{equation}
where we note that ${\rm S}_1$ (containing points $x+h$) and ${\rm S}_2$
(containing points $x$) can in general be sums over different numbers of
streams so that the change in the number
of streams has to be taken into account when performing the limit 
(cf. Fig.~\ref{fig:caustic_limit} and our discussion below); $g_s$ and $g_t$ 
indicate the values of $g$ on the various intersection points of the sheet
with points $x+h$ and $x$ respectively. 

In what follows,
we first discuss the properties of derivatives across such discontinuities
before we turn to the properties of derivatives away from discontinuities
which is almost everywhere (in the mathematical sense) in configuration space.

We wish to remark that this division of space into regions where the derivative
is non-singular (i.e. away from caustics) and regions of singular derivatives
of measure zero is only possible with the proposed explicit phase-space
projection method. Any method that uses (implicit) coarse-graining will
necessarily include the singularities integrated over the coarse-graining
scale in configuration space (see below) although (without coarse-graining)
these regions should not contribute to any volume averages due to their
zero volume measure when no coarse-graining is performed.

\vspace{0.3cm}\noindent {\bf Derivatives across discontinuities:}
As we are concerned here with projections into configuration space of the fine-grained distribution
function, discontinuities in the projected fields are real discontinuities of infinitesimal extent. Consequentially, the 
derivative across a discontinuity is a Dirac-$\delta$ function, i.e. for a velocity
field with left-sided limit $v_l$ and right-sided limit $v_r$ at location $x_0$ --
given generically by $v(x) = \left(v_r - v_l\right)\,\Theta(x-x_0) + v_l$,
where $\Theta$ is the Heaviside $\Theta$-function -- the derivative near $x_0$ is given by 
\begin{equation}
\frac{{\rm d}v}{{\rm d}x} = (v_r - v_l)\, \delta_D(x-x_0).
\end{equation}
For the three dimensional cold distribution functions we are concerned with here, these singular derivatives of finite
measure occupy (one- and) two-dimensional subspaces at the caustic locations.
 If no coarse-graining is performed, they thus 
constitute a subset of configuration space that has volume measure zero and is singular. If coarse-graining
is performed (i.e. the singular surfaces are convolved with a kernel or simply integrated over a finite
volume element), the finite jump $(v_r-v_l)$ is recovered on the scale 
on which coarse-graining is performed. Any approach to determine velocity fields that involves coarse-graining
will thus include contributions from the velocity jump at these discontinuities although the associated volume
is zero. The value of these contributions depends explicitly on the coarse-graining scale.
The physical meaning of the singular derivatives is that they describe the motion of the caustics
themselves.
We discuss below how the differential properties of velocity fields can be obtained in a way
that excludes the contribution of these singular subspaces -- {\em a result that can only be obtained 
with the explicit projection method of the reconstructed fine-grained distribution function that 
we discuss here}.

\vspace{0.3cm}\noindent {\bf Derivatives of projected fields away from discontinuities:}
Away from discontinuities, derivatives of projected properties ought to be performed using
the fine-grained distribution and the explicit derivatives of the projection
operator. In general, this will involve non-trivial commutators. {\em It is
thus not possible to perform a projection operation onto a mesh and then
evaluate derivatives using difference operators on that mesh without implicitly
performing a coarse-graining operation}. Below, we
explicitly calculate some basic differential operators applied to the projected
velocity field. We provide the detailed derivation of the expressions below in appendix~\ref{sec:derivation_divrot}.

\vspace{0.3cm}\noindent {\bf Divergence of the velocity field:} The velocity
divergence naturally already appears at the single-stream level. The divergence
of the mean velocity field is given by 
\begin{equation}
\boldsymbol{\nabla}\cdot \left<\mathbf{v}\right> =
\bigl<
\left(\boldsymbol{\nabla}\log\rho\right) \cdot
\left(\mathbf{v}-\left<\mathbf{v}\right>\right) \bigr> + 
\left<\boldsymbol{\nabla}\cdot\mathbf{v}\right> \label{eq:tet_divergence}
\end{equation} 
arising from the sum of a term that is purely due to the
projection of a multi-stream field (first term), and which reflects alignment
of velocities with density gradients, and the projected single-stream velocity
divergence (second term).
 
\vspace{0.3cm}\noindent {\bf Curl of the velocity field:} As already discussed
above, vorticity vanishes on the fine-grained distribution function and is thus
a property of the projected velocity field alone \citep[see also the discussion
in, e.g.,][on the emergence of vortical flow]{Pichon:1999,Wang:2014} and thus a
purely collective phenomenon of multi-streaming. It is in full generality given
by 
\begin{equation} 
\boldsymbol{\nabla}\times \left<\mathbf{v}\right>  = 
 \bigl<\left(\boldsymbol{\nabla}\log\rho\right)\times\left(\mathbf{v}-\left<\mathbf{v}\right>\right)\bigr>
+ \left< \boldsymbol{\nabla}\times\mathbf{v}\right>,\label{eq:tet_curl}
\end{equation} 
{\em but the second term vanishes for gravitationally generated
velocities}. The form of this equation is reminiscent of the baroclinic term in
the vorticity equation of an ideal fluid where vorticity arises from the
mis-alignment of density and pressure gradients.


\subsection{The case of 1+1 dimensional phase-space}
\label{sec:one_D}
In the case of one spatial and one velocity dimension, the fine-grained
distribution function can be approximated by connecting particles that are
neighbours in the initial conditions through straight lines. 
The mass $m$ of one particle can then be thought of as being
distributed uniformly along the line element so that the single-stream density
at every particle location is simply $\rho=2m/l$, where $l=\Delta
x_{i,i-1}+\Delta x_{i,i+1}$ is the sum of the distances to the neighbouring
particles in Lagrangian space. Single-stream velocities and densities can then
be linearly interpolated in-between particles. The density and
velocity projected into configuration space at an arbitrary location $x$ are readily calculated from the
single-stream values by determining which line-elements intersect $x$ and then
calculate the weighted averages as discussed above.

In Fig.~\ref{fig:velocity_estimators}, we show the mean velocity obtained in
this way for the late stages of plane-wave collapse (red line). We compare it
to the estimate that a Delaunay tessellation approach would give (blue). In
this approach, the phase-space connectivity is not respected and only
configuration space information is used. Particles that are closest in
configuration space are connected by linear elements and velocities are
linearly interpolated along these elements. In multi-stream
regions this leads to a noisy and incorrect projected velocity field if no
coarse-graining is performed.

\begin{figure}
\begin{center}
\includegraphics[width=0.45\textwidth]{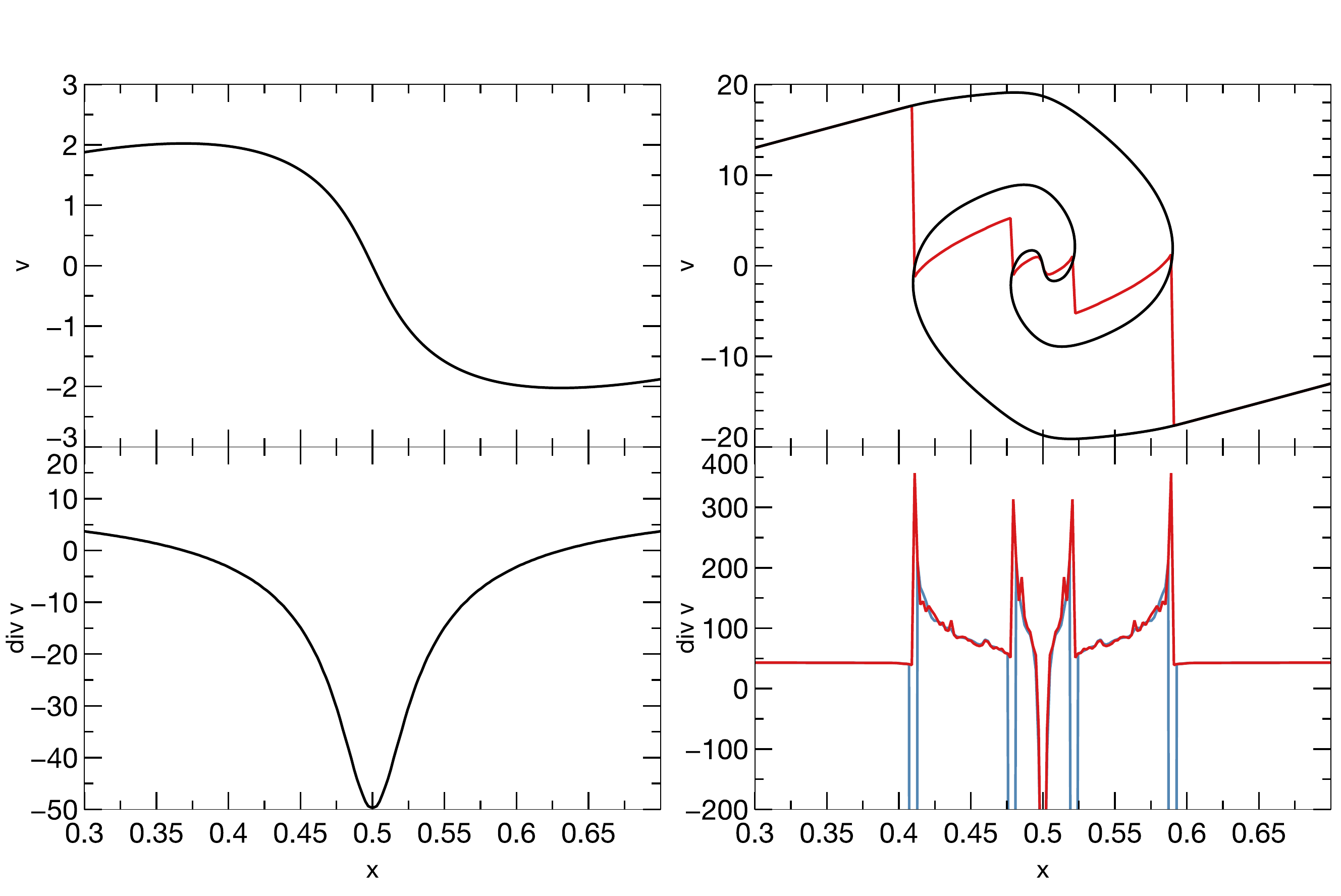}
\end{center}
\caption{\label{fig:sheet_divergence}Before shell-crossing (left column), the
mean velocity field is identical to the fine-grained (top left) and the
velocity divergence (bottom left) is anti-correlated with the overdensity.
After shell-crossing (right column), the mean velocity field (red, top right)
is different from the fine-grained multi-stream velocity field (black). The velocity
divergence (bottom right) estimated directly on the sheet (following
eq.~\ref{eq:tet_divergence}) is everywhere positive except in the very centre.
In contrast, a finite difference estimate computed from gridded values of the
mean field (blue) shows a dominant negative velocity divergence at
caustics but is slightly less noisy. The negative divergence reflects the finite
velocity jump (that is in fact occurring over an infinitesimal volume and is thus 
not present in the direct projection).
The correlation between the velocity
divergence and the overdensity field changes its sign except in the
innermost region and in the single-stream region.  }
\end{figure}

We can see from Fig.~\ref{fig:velocity_estimators} that the mean velocity field
(red line) has discontinuities at the caustics and, at that particular time, also
has a positive slope everywhere except in the very centre. We argued in
the previous subsection that these discontinuities (due to a change in the
number of streams) lead to errors when finite differences are used and the
finite difference interval brackets a caustic. In
Fig.~\ref{fig:sheet_divergence}, we show the phase-space (top panels)
together with the velocity divergence (bottom panels) for the collapse of a
plane wave before (left panels) and after (right panels) shell crossing. Before
shell-crossing, the velocity field is single-valued and has negative velocity
divergence in the central region $0.4\lesssim x\lesssim 0.6$. After
shell-crossing, the mean velocity field is given by the red line (top right
panel) with the properties we have just discussed. If the velocity divergence
is computed by applying finite differences to the projected field on a grid
(blue line, bottom right panel), the singular derivative (see discussion above)
at the caustic locations
is effectively integrated over a finite volume so that the finite velocity jump divided
by the scale is recovered. 
If eq.~(\ref{eq:tet_divergence}) is used to compute the velocity divergence (red
line, bottom right panel), i.e. by computing single stream finite differences
on the phase space line elements followed by projection onto configuration
space, the singular derivative, which exists only at a finite number of points
need not be included. As a result, we see that the velocity field is expansive
almost everywhere with discontinuous jumps at the caustics.

We observe that the divergence calculated directly from the sheet in the 
described way is somewhat noisy. This is a consequence of the
low-order interpolations we are using. Higher order schemes are expected to
improve this but are beyond the scope of this article.

Another interesting observation from Fig.~\ref{fig:sheet_divergence} is that,
before shell crossing, the velocity divergence is anti-correlated with
overdensity -- divergence being negative where overdensity is positive -- while
after shell-crossing this correlation reverses except in the very centre of the
collapsed structure where the divergence oscillates between positive and
negative. In the 1+1 dimensional case, the velocity divergence of collapsed
structures is thus predominantly positive. We will revisit this aspect in the
3+3 dimensional case in Section~\ref{sec:velocity_slices}.


\subsection{Projections of phase space in $N$-body simulations and the dark matter sheet}

In our discussion above, we have argued that knowledge of the distribution
function is {\em necessary} to accurately perform its projection onto
configuration space and thus determine the mean velocity field {\em without 
performing a coarse-graining operation}. As we have
discussed in AHK12, the fine-grained distribution function of cold dark matter
can be reconstructed from $N$-body simulations using the phase-space sheet
tessellation method described there. In this method, a tessellation of the
Lagrangian particle coordinates $\mathbf{q}$ is performed, decomposing the
entire particle distribution into a collection of tetrahedral phase space
elements. 

Since, due to Liouville's theorem, the tessellation is preserved in phase
space, it can be reconstructed at any later time if the connectivity is known.
In the case of an initial cubical lattice of particles, this is particularly
simple since the three-dimensional Lagrangian coordinate can be encoded in the
particle IDs, and the connectivity can be determined on-the-fly. In fact, the
entire procedure to know which four particles span every relevant simplex
(tetrahedron) of the tessellation reduces to decomposing one unit cube into six
tetrahedra of equal size (the Delaunay triangulation of the unit cube).
Consequently, one can then think of the data output from a simulation as the
information that describes the evolution of the vertices of the $N_{tet}\equiv
6 N_p$ tetrahedra. Each of them carries $m_{\rm tet}=M_{\rm box}/N_{\rm tet}$
of the mass $M_{\rm box}$ in the simulation box and represents a piecewise
linear interpolation between four vertices on the fine-grained distribution
function. In fact, to reduce the possible impact of anisotropies due to the choice
of one of the six equivalent Delaunay triangulations of the unit cube, we
average over all those six possibilities. Similar methods employing the
advection of a tessellation of test particles can be used in simulations of
incompressible flows, allowing the calculation of Lagrangian flow properties
\citep[e.g.][]{Pumir:2013}.


\subsubsection{Interpolating and differentiating on tetrahedra}
\label{sec:app_finitediff}
For a tetrahedron with vertices $\mathbf{x}_k = (x_k,y_k,z_k)$, $k=1\dots4$, any point $\mathbf{x}=(x,y,z)=(x_i)$, $i=1\dots3$, can be expressed in terms of {\it tetrahedral coordinates} $\boldsymbol{\zeta}=(\zeta_1,\zeta_2,\zeta_3,\zeta_4)$ through the linear transformation 
\begin{equation}
\left( 1,x,y,z \right)^{T}   = \left[ \begin{array}{cccc}
1 & 1 & 1 & 1 \\
x_1 & x_2 & x_3 & x_4 \\
y_1 & y_2 & y_3 & y_4 \\
z_1 & z_2 & z_3 & z_4 
\end{array}\right] \cdot \boldsymbol{\zeta}^{T} =: {\rm J} \cdot \boldsymbol{\zeta}^{T}.
\end{equation}
It is easy to see that $\sum_k \zeta_k = 1$ and the volume of the tetrahedron is simply $V =  \frac{1}{6}\left| \det\,{\rm J} \right|$. For any field $F$ whose values are known at the vertices $\mathbf{x}_k$, i.e. $F_k=F(\mathbf{x}_k)$, the linear interpolation to an arbitrary point $\mathbf{x}$ can then be written as
\begin{equation}
F\left[\boldsymbol{\zeta}\left(\mathbf{x}\right)\right] = \sum_{k=1\dots4} F_k \,\zeta_k(\mathbf{x}). \label{eq:lin_interp}
\end{equation}
This implies that differentials of $F$ can be obtained through the chain-rule by simply computing
\begin{equation}
\frac{\partial F}{\partial x_i} = \sum_{k=1\dots4} \frac{\partial F}{\partial \zeta_k} \frac{\partial \zeta_k}{\partial x_i} = \sum_{k=1\dots4} {\rm J}^{-1}_{i+1,k} \, F_k,
\label{eq:tet_deriv}
\end{equation}
where the index $i$ refers to a Cartesian coordinate, and {\it not} to vertex $i$.
Note that a derivative computed in this way is only accurate at first order (like the simple backward/forward finite difference operators). Gradient operators that are accurate at higher order are of course possible and would include information from more than one tetrahedron or involve quadratic
tetrahedra (defined using 10 instead of 4 vertices for the linear tetrahedron). We leave this aspect for future work as first order gradients are accurate enough for the purposes of this article. We note however that, for example,  a curl-free non-linear field will appear to have curl if evaluated with low-order finite differences. We note that differentials only exist if ${\rm J}$ is invertible, i.e. if $\det {\rm J}\neq 0$, which is equivalent to $V\neq 0$. 


\subsubsection{Determining the fine-grained density and velocity as well as their derivatives}
The sheet is parametrised by the Lagrangian coordinate $\mathbf{q}$ and discretised into tetrahedral elements with vertices $\mathbf{q}_{i}$. If a piecewise constant density field is sufficient, one can simply use the location of the vertices (particles) to determine the mass density that every tetrahedron contributes to a point inside its volume as $\rho_{\rm tet}=m_{\rm tet}/V_{\rm tet}$, with the tetrahedron volume $V_{\rm tet}$ defined above.
As we have discussed in Section~\ref{sec:derivatives} above, the derivatives of the projected velocity field contain also derivatives of the fine-grained density. Hence, a piecewise constant fine-grained density will not be sufficient. We therefore compute a density estimate at each vertex from the mean volume of all $N_{\rm tets}$ tetrahedra that share a given vertex $\mathbf{q}_i$. Specifically, the density at a given vertex $\mathbf{q}_i$ is given by $\rho(\mathbf{q}_i) = m_{\rm tet}N_{\rm tets}\left[\sum_{j=1\dots N_{\rm tets}(\mathbf{q}_i)}V_j\right]^{-1}$, i.e. the reciprocal of the sum over the volumes $V_j$ of all tetrahedra that share the vertex. We note that it is not possible to use the dual Voronoi mesh to obtain a vertex-centered density since the Voronoi cells are not necessarily convex. 

Having a density estimate defined at each vertex, we can now linearly interpolate easily to any position $\mathbf{x}$ inside a given tetrahedron using eq.~(\ref{eq:lin_interp}). This achieves a piecewise linear field with a piecewise constant derivative across the tessellation. Similarly, for each tetrahedron, the fine-grained velocity $\mathbf{v}_{1\dots4}$ is known at the four vertices $\mathbf{q}_i$ and can be linearly interpolated to position $\mathbf{x}$ using eq.~(\ref{eq:lin_interp}), and eq.~(\ref{eq:tet_deriv}) can be used to compute derivatives on each tetrahedron. In this way, the gradient of the fine-grained (i.e. per stream) density field $\boldsymbol{\nabla}\rho$ as well as the gradient tensor of the fine-grained velocity field $\boldsymbol{\nabla}\otimes\mathbf{v}$ can be easily calculated from the values of the density as well as the velocity at the four vertices of each tetrahedron. 


\subsubsection{Projected fields using the dark matter sheet}
To determine the projection of a field $g(\mathbf{q})$, defined on the dark matter sheet, at an arbitrary point $\mathbf{x}$, we first determine all the tetrahedra which contain $\mathbf{x}$. The velocity and density due to each stream is computed by linearly interpolating from the four vertices of each tetrahedron to the point. Having now a linearly interpolated quantity $g_i(\mathbf{x})$ for tetrahedron $i$, we next average over all the tetrahedra $i$ that contain that point and weight each contribution by the density of the tetrahedron, i.e. make the approximation
\begin{equation}
\left<g\right>(\mathbf{x})\simeq\frac{\sum_i^{N_{t}} \rho_i(\mathbf{x})\,g_i(\mathbf{x})}{\sum_i^{N_{t}} \rho_i(\mathbf{x})},
\end{equation}
where $N_{t}$ is the number of tetrahedra that intersect point $\mathbf{x}$. Note that $g_i$ is only defined inside the corresponding tetrahedron $i$ and is equal to zero outside. This is done most efficiently by looping over all tetrahedra and adding their contributions to all the cells in the uniform grid whose cell centres are contained inside the tetrahedron. In fact, we subsample each cell 8 times, i.e. we compute the value for each cell by averaging over the values at 8 points inside the cell to arrive at a value closer to an actual volume average for each cell. We perform this operation for all points of a cubical lattice to obtain a three-dimensional data cube of velocity information.



\section{Simulations}
\label{sec:sims}
We have performed a series of cosmological simulations covering a considerable range in box sizes and mass resolutions to ensure convergence of our results. In all cases, we generated initial conditions using {\sc music} \citep{Hahn:2011} adopting the parametrisation of the  transfer function of \cite{Eisenstein:1999} for the cold-dark matter (CDM) simulations as well as a truncated transfer function for the warm-dark matter (WDM) simulations that we detail below. We use cosmological parameters consistent with the WMAP7 data release \citep{Komatsu:2011}. Specifically, we use the density parameters $\Omega_m = 0.276$, $\Omega_{\Lambda} = 0.724$ and $\Omega_{b} = 0.045$, a Hubble parameter of $h = 0.703$, power spectrum normalisation $\sigma_8 = 0.811$ and spectral index $n_s = 0.96$. 
 
The initial conditions are generated by perturbing a regular lattice of particles using the Zel'dovich approximation at $z=100$.  The details on all the simulations that we use in this work are summarised in Table~\ref{tab:sims}. All our analysis is performed at $z=0$.
 
 
 \subsection{CDM simulations}
For the CDM simulations, our four box sizes range between $3\,h^{-1}{\rm Gpc}$ and $100\,h^{-1}{\rm Mpc}$, the mass resolution is $512^3$ particles in all cases, the $1\,h^{-1}{\rm Gpc}$ box has been also run at a higher resolution of $1024^3$ particles. The gravitational evolution between $z=100$ and $z=0$ for these simulations has been performed using the tree-PM code {\sc L-Gadget 3} \citep{Angulo:2012a} using a $1024^3$ PM mesh for the simulations with $512^3$ particles and a $2048^3$ mesh for the simulations with $1024^3$ particles. The softening adopted for the tree force is given in Table~\ref{tab:sims} for all cases.


\subsection{WDM simulations}
We consider also simulations starting from a perturbation spectrum with small-scale suppression in this work. This has the advantage that, unlike in CDM, the simulations are able to capture the full dynamic range of perturbations with enough resolution. We consider the same simulations as in \cite{Hahn:2012}, i.e., specifically, we adopt the parametrisation of the WDM transfer function from \cite{Bode:2001} to modify the CDM transfer function
\begin{equation}
 T_{\rm WDM}(k) = T_{\rm CDM}(k) \left[1 + (\alpha\,k)^2\right]^{-5.0},
\label{Tk-WDM}
\end{equation}
and
\begin{equation}
\frac{\alpha}{h^{-1}{\rm Mpc}}  \equiv 0.05 \left(\frac{\Omega_m}{0.4}\right)^{0.15} \left(\frac{h}{0.65}\right)^{1.3}
\left( \frac{m_{\rm dm}}{1\,{\rm keV}}  \right)^{-1.15},
\label{alpha}
\end{equation}
which for the WDM particle mass of $300\,{\rm eV}$ equals a cut-off scale $\alpha=0.21\,h^{-1}{\rm Mpc}$. For these simulations, we employed a modified version of {\sc Gadget-2} \citep{Springel:2005} that uses only the particle-mesh force, evaluated on a $512^3$ mesh (Note that we employ only the standard $N$-body simulations from \citealt{Hahn:2012} here). We simulated the gravitational evolution of one $40\,h^{-1}{\rm Mpc}$ box using $128^3$, $256^3$ as well as $512^3$ particles. The cut-off scale $\alpha$ is resolved in all three runs.

\begin{table}
\begin{center}
\begin{tabular}{lcccc}
Simulation  & $L_{\rm box}$ & $N_{\rm p}$ & $m_{\rm p}$ & $\epsilon$\\
Name & $h^{-1}{\rm Mpc}$ & & $h^{-1}{\rm M_\odot}$ & $h^{-1}{\rm kpc}$\\
\hline
L3000N512 & $3000$ & $512^3$ & $1.5\times10^{13}$ & 200\\
L1000N512 & $1000$ & $512^3$ & $5.7\times10^{11}$ & 65\\
L300N512 & $300$ & $512^3$ & $1.5\times10^{10}$ & 20\\
L100N512 & $100$ & $512^3$ & $5.7\times10^{8}$ & 6.5\\
L1000N1024 & $1000$ & $1024^3$ & $7.1\times10^{10}$ & 35\\
\hline
WDM512 & $40$ & $512^3$ & $3.7\times10^{7}$ & 78 \\
WDM256 & $40$ & $256^3$ & $2.9\times10^{8}$ & 78 \\
WDM128 & $40$ & $128^3$ & $2.3\times10^{9}$ & 78 \\
\hline
\end{tabular}
\end{center}
\caption{\label{tab:sims}Labels and specifics of the simulations used in this work.}
\end{table}


\section{Real-space properties of the cosmic velocity field}
\label{sec:realspaceresults}
In this section, we present the results of our analysis of velocity fields in real space. First, we show slices of the velocity divergence and vorticity in cosmological simulations of CDM and WDM structure formation. We then present their 1-point statistical properties and give a preliminary discussion of the velocity fields inside dark matter haloes.

\subsection{Validation}

 \begin{figure}
\begin{center}
\includegraphics[width=0.4\textwidth]{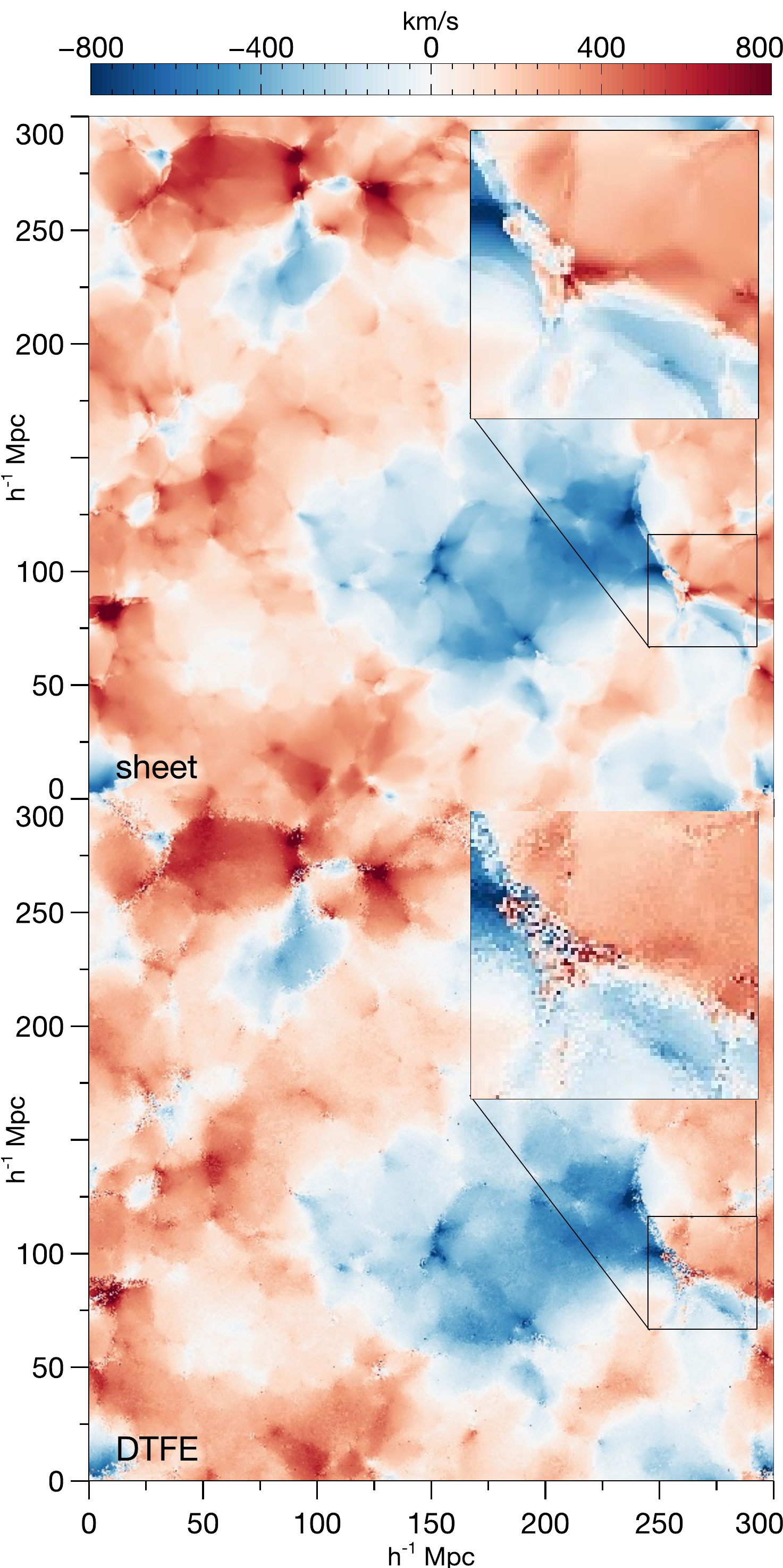}
\end{center}
\caption{\label{fig:velocity_panels}Slice of the cosmic velocity field from the L300N512 simulation obtained with the phase space tessellation method discussed in this work (top) and the DTFE method (bottom). Shown is the $x$-component of the velocity. The insets show enlargements of a $50\times50\,h^{-2}{\rm Mpc}^2$ region highlighting differences between the two methods in multi-stream regions. Equivalently to Fig.~\ref{fig:velocity_estimators}, noise in the lower panel inset arises  due to interpolation between unrelated particles in the absence of phase-space information. Obtaining the correct point-wise mean-velocity field in multi-stream regions is only possible through a projection of the reconstructed fine-grained distribution function.
}
\end{figure}

 \begin{figure}
\begin{center}
\includegraphics[width=0.4\textwidth]{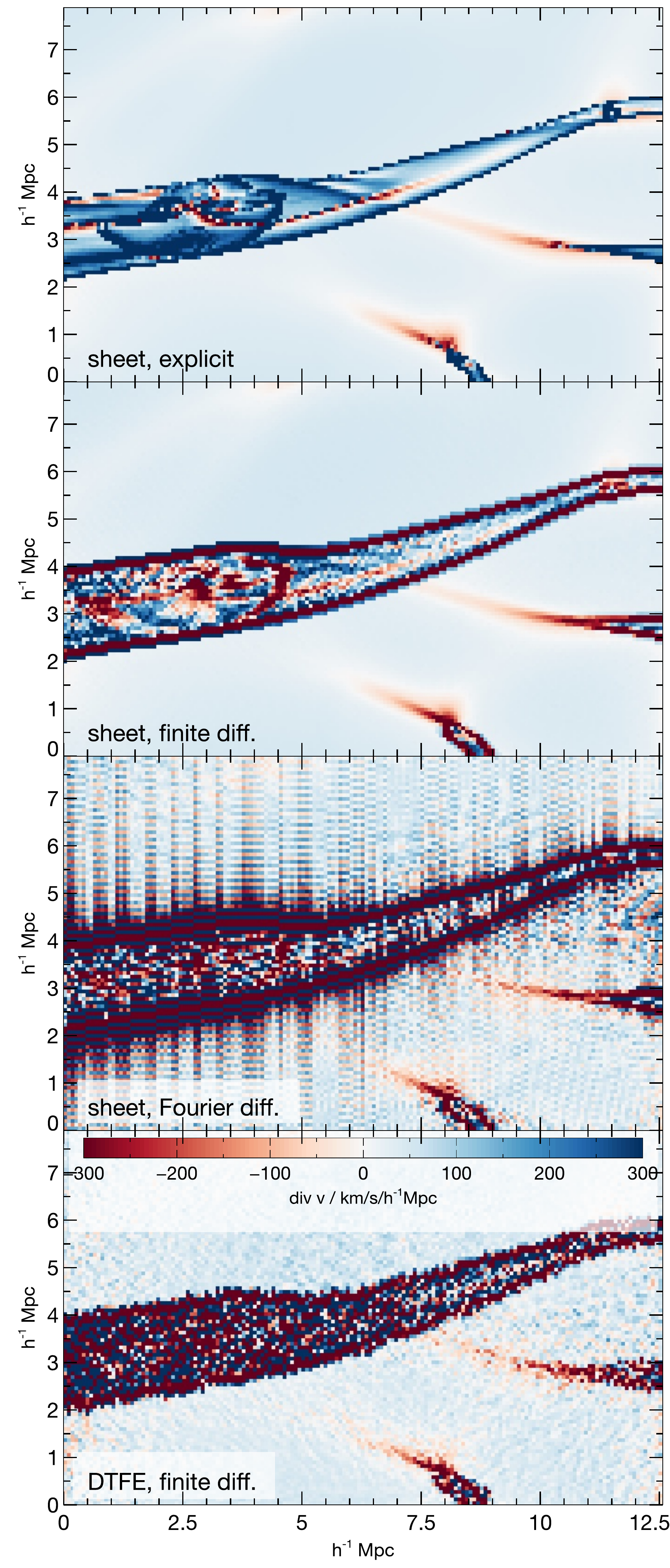}
\end{center}
\caption{\label{fig:divv_panels}Slice of the velocity divergence field through a filament in the WDM512 simulation. We show results obtained with velocity fields computed using the dark matter sheet (top three panels) and using DTFE (bottom panel). The divergence is computed using the explicit divergence of the projected field from eq.~(\ref{eq:tet_divergence}) (top), 4th order finite differences on a mesh (second from top and bottom) as well as using spectral derivatives (third from top). See text for details.
}
\end{figure}

We compare our results for velocity fields with corresponding results obtained
with a Delaunay tessellation in Eulerian space, i.e. based on the particle
positions only, which has been employed commonly in the literature
\citep[e.g.][]{Icke:1987, Bernardeau:1996,Pueblas:2009,Jennings:2011}. Such a
method does not involve a projection of the actual distribution function, but
interpolates between the particles on the fine-grained distribution in
configuration space (i.e. performs the tessellation after the projection) which
necessarily leads to inconsistencies in multi-stream regions when the mesh
on which the velocities are evaluated is of higher resolution than the
Delaunay tessellation. We expect, however, this approach to perform better
when the mesh cells are large and an effective coarse-graining is 
performed. As in our
approach, velocities at an arbitrary point $\mathbf{x}$ can be determined by
linearly interpolating the velocity of the four vertices (particles) of the
Delaunay tetrahedron containing $\mathbf{x}$ to $\mathbf{x}$. To perform this
calculation, we used the publicly available version 1.1.1 of {\sc
dtfe}\footnote{\tt{http://www.astro.rug.nl/$\sim$voronoi/DTFE/dtfe.html}}
\citep{Cautun:2011}.

We have already discussed the difference between employing only configuration space
information (such as DTFE) and using full phase space information in Section~\ref{sec:one_D}.
Similar differences become readily visible when we inspect
slices of the velocity field in one of the cosmological simulations. In the top
panel of Fig.~\ref{fig:velocity_panels}, we show the $x$-component of the
velocity field for the L300N512 run. In the bottom panel, we show the
respective velocity field obtained using DTFE. In both cases, we have resampled
to a cube of $512^3$ cells. For the Delaunay tessellation, the small scale
jitter due to linear interpolation between vertices that are not close in phase
space and the lack of averaging in multi-stream regions is clearly apparent in
the insets that zoom into the velocity field.

As discussed in Section~\ref{sec:derivatives}, it is non-trivial to compute
derivatives of the mean velocity field since a commutator between the
derivative and the projection operator appears. We will illustrate this now by
looking at slices of the velocity divergence field computed from the WDM512
simulation in Fig.~\ref{fig:divv_panels} using various derivative estimators.
Specifically, we compare the correct expression based on the fine-grained
distribution function from eq.~(\ref{eq:tet_divergence}) (top left panel), with
a divergence computed on the gridded velocity field using a 4th order finite
difference divergence operator (top right) and with the Fourier-space
divergence operator (bottom left) that we will later employ when computing the
spectral properties of velocity fields. The latter is given by computing ${\rm
div }\,\mathbf{v} = \mathcal{F}^{-1}\left[-i\mathbf{k}\cdot
\mathcal{F}\left[\mathbf{v}\right]\right]$ using the fast Fourier transform
(FFT) on the mesh on which we have computed the velocity field. Finally, we
also show results when computing the velocity divergence for the DTFE estimate
using again the 4th order finite difference operator. The differences for the
velocity divergence fields computed using the various methods are readily
visible. As expected, computing derivatives on the noisier DTFE field leads to a
noisy divergence field. The spectral derivative performs badly by
producing Gibbs ringing due to the discontinuous field. Finally, we also
clearly see that the finite difference estimate performs an implicit coarse-graining
by differentiating on a finite scale across caustics leading to strongly
compressive features on the infall side of every caustic, just as in the
one-dimensional case that we have discussed above in
Section~\ref{sec:coarse_grained_vel}. In fact, these compressive features,
while depending on the scale on which the derivative is taken, are
clearly the highest magnitudes of divergence compared to what is estimated 
away from caustics with eq.~(\ref{eq:tet_divergence}).

By using phase-space information to average over multi-stream regions, we thus
expect that our method will greatly improve estimates of the true discontinuous nature of the
velocity field and its derivatives on non-linear scales. With our approach it is
possible to determine derivatives on significantly smaller scales than before, in fact even point-wise,
without increasing the noise. We explore these aspects in more detail in the
remainder of this paper.


\begin{figure*}
\begin{center}
\includegraphics[width=0.9\textwidth]{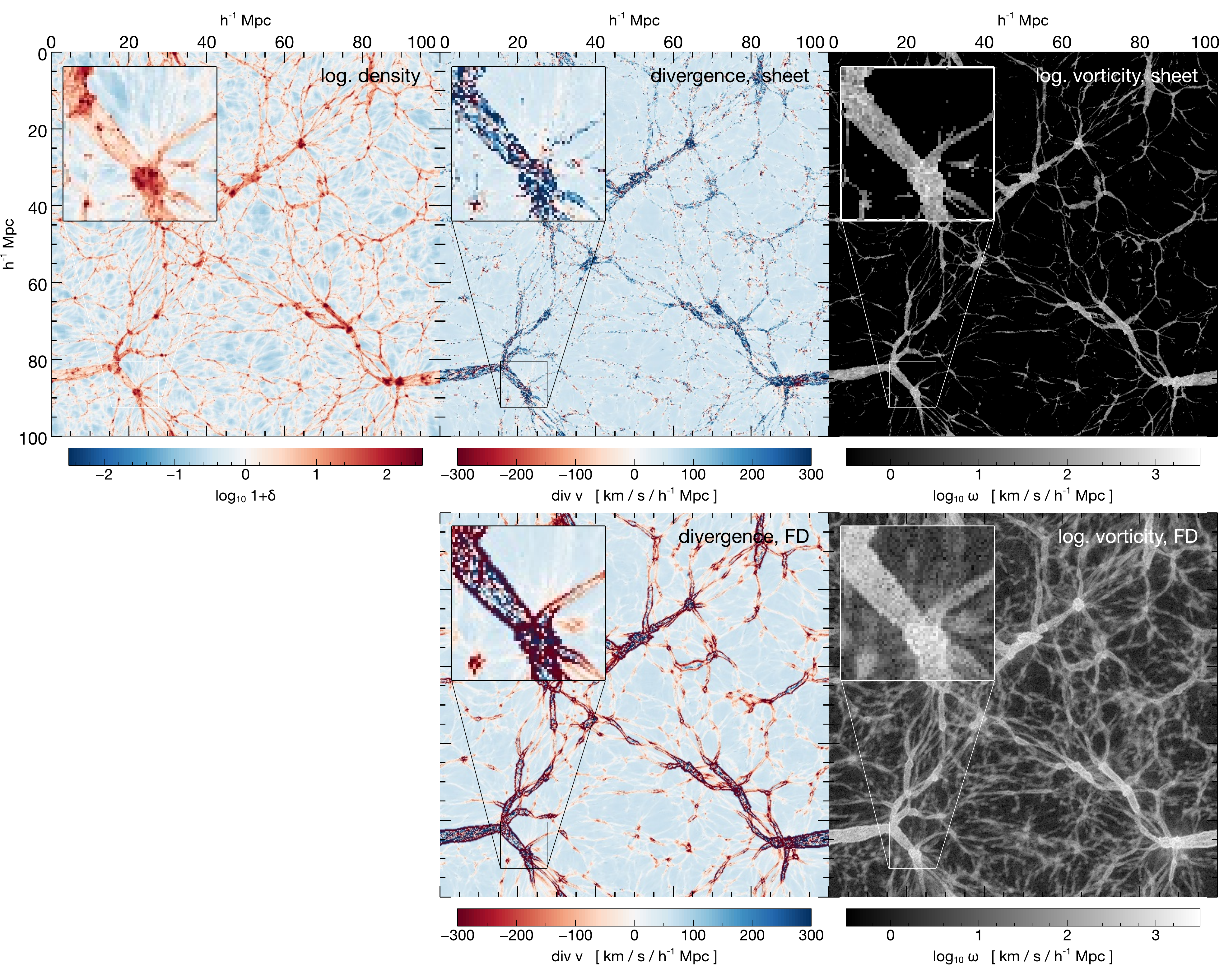}
\end{center}
\caption{\label{fig:velocity_props}Slice through the overdensity (left panel), velocity divergence (middle panels) and the vorticity (right panels) fields of the L100N512 simulation. Divergence and vorticity in the top row have been calculated directly from the dark matter sheet using eqs.~(\ref{eq:tet_divergence}) and (\ref{eq:tet_curl}), while those in the bottom row have been calculated from velocity fields computed from the dark matter sheet for which divergence and vorticity were calculated using finite differences. Note that the vorticity is plotted logarithmically, while the velocity divergence (due to its changing sign) is not. 
}
\end{figure*}

\subsection{Divergence and vorticity in CDM cosmic velocity fields}
\label{sec:velocity_slices}
As we have discussed above and shown already in Fig.~\ref{fig:divv_panels},
pronounced differences in the real-space properties exist between divergence
and vorticity estimated directly from the sheet and estimates based on finite
differencing of the discontinuous velocity field. 

In Fig.~\ref{fig:velocity_props}, we show slices through the CDM simulation
L100N512, our simulation resolving the smallest scales. Specifically, we show
the overdensity field (left panel), the velocity divergence (middle panels) as
well as the vorticity field (right panels). Note that the colour map of the
velocity divergence has been inverted with respect to that of the density field
(also in the remainder of the paper), to allow for a more easy comparison in
terms of the anti-correlation between the two fields in linear theory. Again,
we show two versions of velocity derivative estimation, the sheet-based
estimate following eqs.~(\ref{eq:tet_divergence}) and (\ref{eq:tet_curl}) in
the top panels as well as a finite difference estimate in the bottom panels.
All fields were sampled onto a cubical lattice of $512^3$ cells and each slice
corresponds to one slice of the data cube.

We recover for these CDM simulations the same qualitative differences in the velocity divergence fields as already discussed above for the plane-wave as well as the WDM case: finite differencing leads to filaments surrounded by envelopes of strong compression which are a result of the differentiation across the caustic on a finite scale (see Section~\ref{sec:derivatives}). In the sheet-based estimates, the velocity field is almost everywhere expansive, aside from small filaments that have not shell-crossed yet. This results is not easily discernible using finite-difference estimates for the derivatives. 

Similar differences resulting from the derivative estimation can be seen for the vorticity field. Applying eq.~(\ref{eq:tet_curl}) leads to a vorticity  that is explicitly zero in single-stream regions. The finite difference approach fails to recover this aspect. In addition, due to differentiation across caustics, a similar envelope of high vorticity is visible at caustics as the maxima of compression discussed above. We observe that, as discussed by \cite{Pichon:1999}, a jump in vorticity occurs at the same locations as the caustics and with a similar magnitude in velocity change.

\begin{figure*}
\begin{center}
\includegraphics[width=0.65\textwidth]{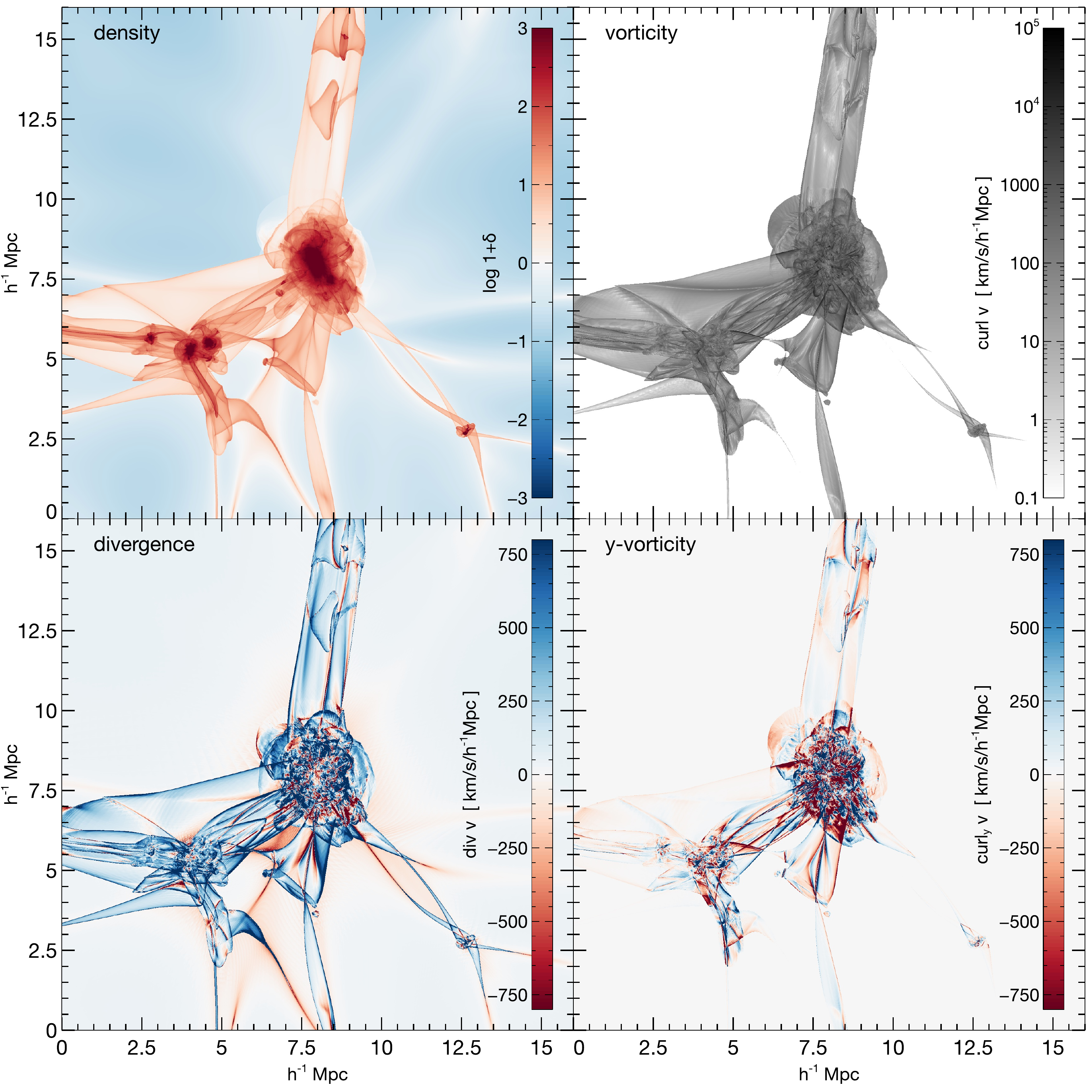}
\end{center}
\caption{\label{fig:halo_slices}Slices through the centre of the most massive halo from the WDM512 simulation. {\bf Top left:} overdensity field in logarithmic scale; {\bf bottom left:} velocity divergence field, convergent flow is shown in blue, divergent in red; {\bf top right:} the vorticity field in logarithmic scale; {\bf bottom right:} $y$-component of the vorticity vector field in linear scale. The extent of each slice corresponds to $16\,h^{-1}{\rm Mpc}$, the slice itself is infinitesimally thin.
}
\end{figure*}

\begin{figure}
\begin{center}
\includegraphics[width=0.42\textwidth]{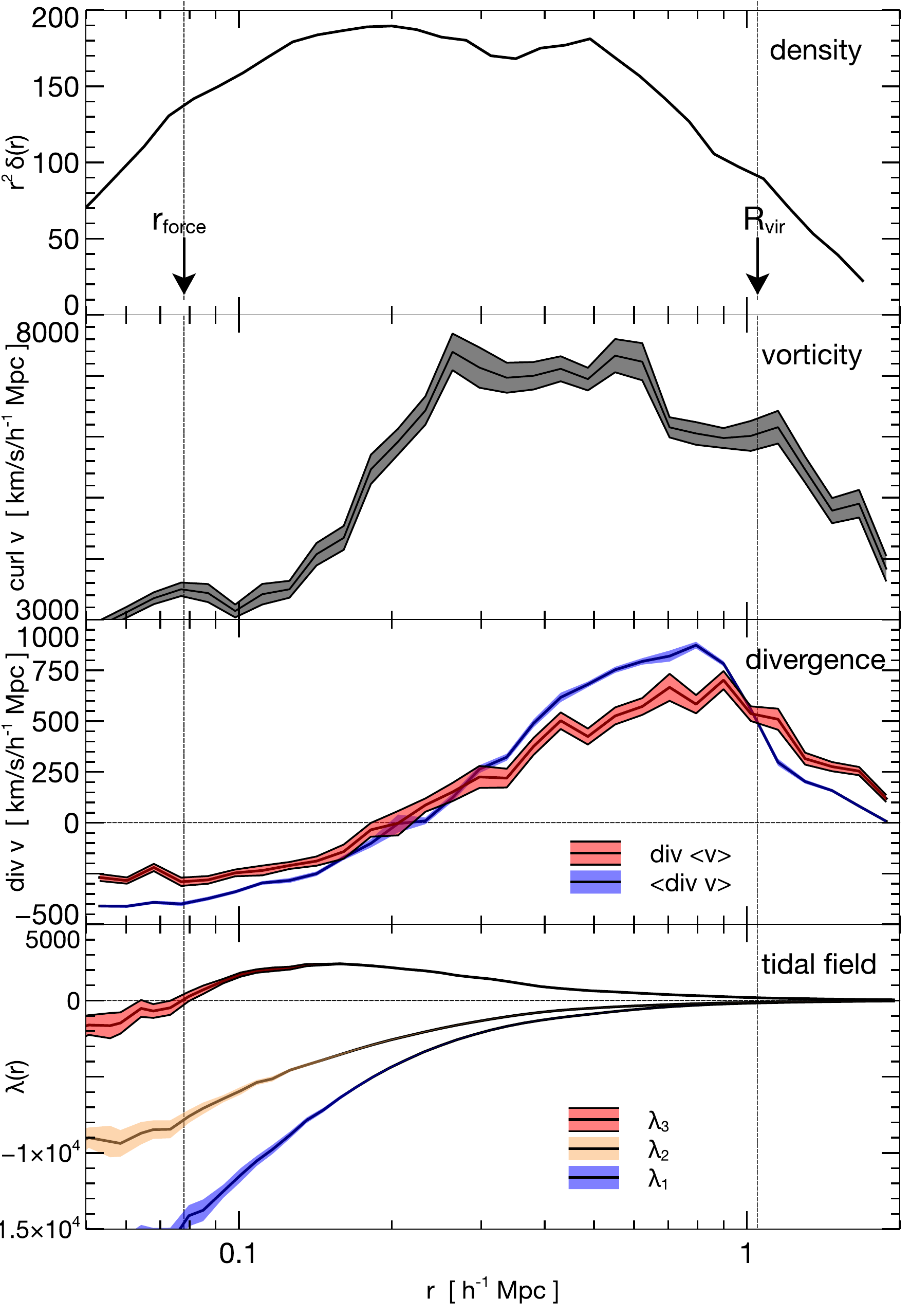}
\end{center}
\caption{\label{fig:halo_profiles}Halo profiles for the most massive halo in
WDM512. Panels are, from top to bottom: the radial overdensity profile;
the vorticity of the mean velocity field;  divergence of the mean velocity field
(red) and the average velocity divergence of single streams (blue); eigenvalues
of the tidal tensor. Shaded regions show the median as well as the error on the
median based on the 16th and 84th percentile. A convergence study of the
velocity divergence is shown in Appendix~\ref{sec:resolution}.  }
\end{figure}

\begin{figure}
\begin{center}
\includegraphics[width=0.44\textwidth]{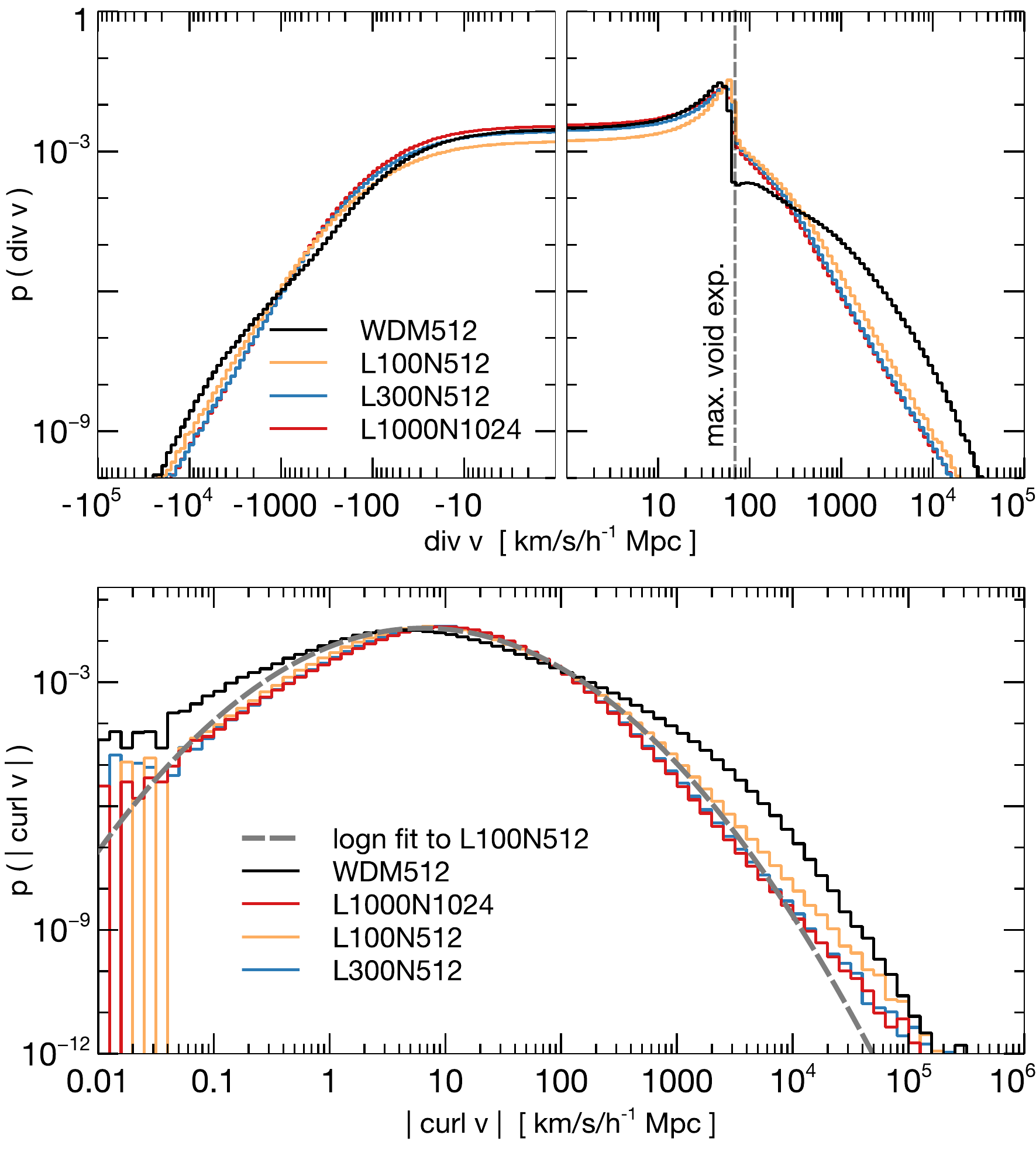}
\end{center}
\caption{\label{fig:velocity_pdfs}Volume weighted probability distribution functions of the velocity divergence (top) and vorticity (bottom) for various simulations. The vertical dashed black line in the top right panel indicates the maximum velocity divergence for a completely empty void (i.e. $\delta=-1$, for which ${\rm div}\,\mathbf{v} = \frac{3}{2}\Omega_m^{0.6}H_0$). The dashed gray line in the bottom panel shows the reasonably good fit of a lognormal distribution to the vorticity PDF of simulation L100N512.
}
\end{figure}


\subsection{The velocity field in haloes}
\label{sec:single_halo}
We continue our qualitative discussion of velocity fields by investigating in more detail the most massive halo of mass $M_h\sim1.4\times10^{14}\,h^{-1}{\rm M}_\odot$ (which corresponds to $\sim4$~million particles) in our WDM512 simulation. We perform this analysis for the WDM case since the truncation of the perturbation spectrum allows us to resolve the collapse of those perturbations that are present with enough resolution. This cannot be achieved in the CDM case since perturbations up to the resolution limit exist.

In Fig.~\ref{fig:halo_slices}, we show slices of the density field (top
left), velocity divergence (bottom left), the vorticity (top right) as well as the
$y$-component of the vorticity vector field (bottom right) through the centre of this most
massive halo. We now clearly see how the velocity divergence is positive in
both void regions as well as predominantly positive in regions that have shell
crossed. The only regions of {\em con}vergence are given by regions of moderate
overdensity that are not yet shell-crossed (but will do so in the future and
are visible in red colour in the divergence map and not visible in the
vorticity magnitude maps in Fig.~\ref{fig:halo_slices}). This can be clearly seen from
the vorticity slice, where the convergent regions are not visible, indicating
that they have not yet shell-crossed. The vorticity increases with every caustic,
being highest in the outer parts of the virialised regions of the halo. In the
very centre of the halo, the velocity field is convergent and the vorticity
decreases again. 

We show these aspects in more details in Fig.~\ref{fig:halo_profiles}, where
we plot spherically averaged profiles of the density, vorticity, velocity
divergence and the three tidal field eigenvalues. The density profile has not
been calculated with the dark matter sheet, instead we simply plot the
overdensity of $N$-body particles per shell to avoid the centrally biased
profiles we have observed for the sheet in AHK12. The drop in vorticity and,
more importantly, the sign reversal of the velocity divergence in the central
region are now more clearly visible than in Fig.~\ref{fig:halo_slices}.

We observe a behaviour of the vorticity and divergence profiles that is
consistent with our visual inspection of the slices. Velocity divergence rises
as we enter the virial radius, is highest just inside the virial radius, then
declines and becomes negative for $r\lesssim 0.2\,h^{-1}{\rm Mpc}$. This
qualitative behaviour is present for both the divergence of the mean velocity
field, as well as the averaged single stream divergence. We note that, at least
for the halo we inspect, the radius where the divergence flips sign is close to
the point where $\frac{{\rm d}\log \delta}{{\rm d}\log r}\sim2$.  In
Appendix~\ref{sec:resolution}, we show that while the details of the velocity
divergence profiles converge rather slowly with increasing resolution, the
qualitative features -- i.e. the shape of the profile as well as the peak
position and, most importantly, the radius at which the profile switches sign
-- are remarkably stable across resolutions. 

One might thus speculate whether the change from convergent to divergent could
be caused by a corresponding change of the tidal field. We also show the three
eigenvalues of the tidal tensor in Fig.\ref{fig:halo_profiles}. The tidal
field eigenvalues $\lambda_1\leq\lambda_2\leq\lambda_3$ are obtained by
diagonalising the tidal field tensor ${\rm
T}_{ij}\equiv-\partial_i\partial_j\widehat{\phi}$, where $\widehat\phi$ is the
gravitational potential normalised in such a way that ${\rm tr\,}{\rm T}_{ij} =
\lambda_1+\lambda_2+\lambda_3 = -\delta$. We find that the eigenvalue signature
is $(--+)$ at all radii, implying one-dimensional stretching and
two-dimensional compression. It thus seems not very plausible that the change
from convergent to divergent flow is of a simple tidal origin. The associated
drop in vorticity might point towards an isotropised inner region, where
particle velocities are well aligned with density gradients. We only pointed
out a few, rather qualitative, observations of the properties of mean velocity
fields in haloes. A more rigorous and detailed inspection is beyond the scope
of this paper and will be followed up in future work.


\subsection{One-point statistics}

In this section, we return to the global statistics of the velocity field in
cosmological volumes. In Fig.~\ref{fig:velocity_pdfs}, we show the volume
weighted probability distribution functions (PDFs) of the velocity divergence
and vorticity fields. We find that the CDM vorticity PDFs that we show are
reasonably well described by a lognormal distribution 
\begin{equation}
p\left(\omega \left|\mu,\sigma\right.\right) = \frac{1}{\sqrt{2\pi}\sigma\omega}\exp\left(\frac{-\left(\log \omega - \mu\right)^2}{2\sigma^2}\right),
\end{equation}
with best fit parameters $(\mu,\sigma)$ of $(3.55,1.3)$, $(3.42,1.14)$ and
$(3.45,1.10)$ for L100N512, L300N512 and L1000N1024, respectively. Quite in
contrast, the WDM vorticity distribution has a distinctly different shape with
strongly enhanced wings and a change in slope at large vorticity. Similar
behaviour can be observed in the PDFs of the velocity divergence. While the CDM
distributions have power-law tails at both positively and negatively large
values, the WDM distribution has a distinctly enhanced tail. 

\begin{figure*}
\begin{center}
\includegraphics[width=0.4\textwidth]{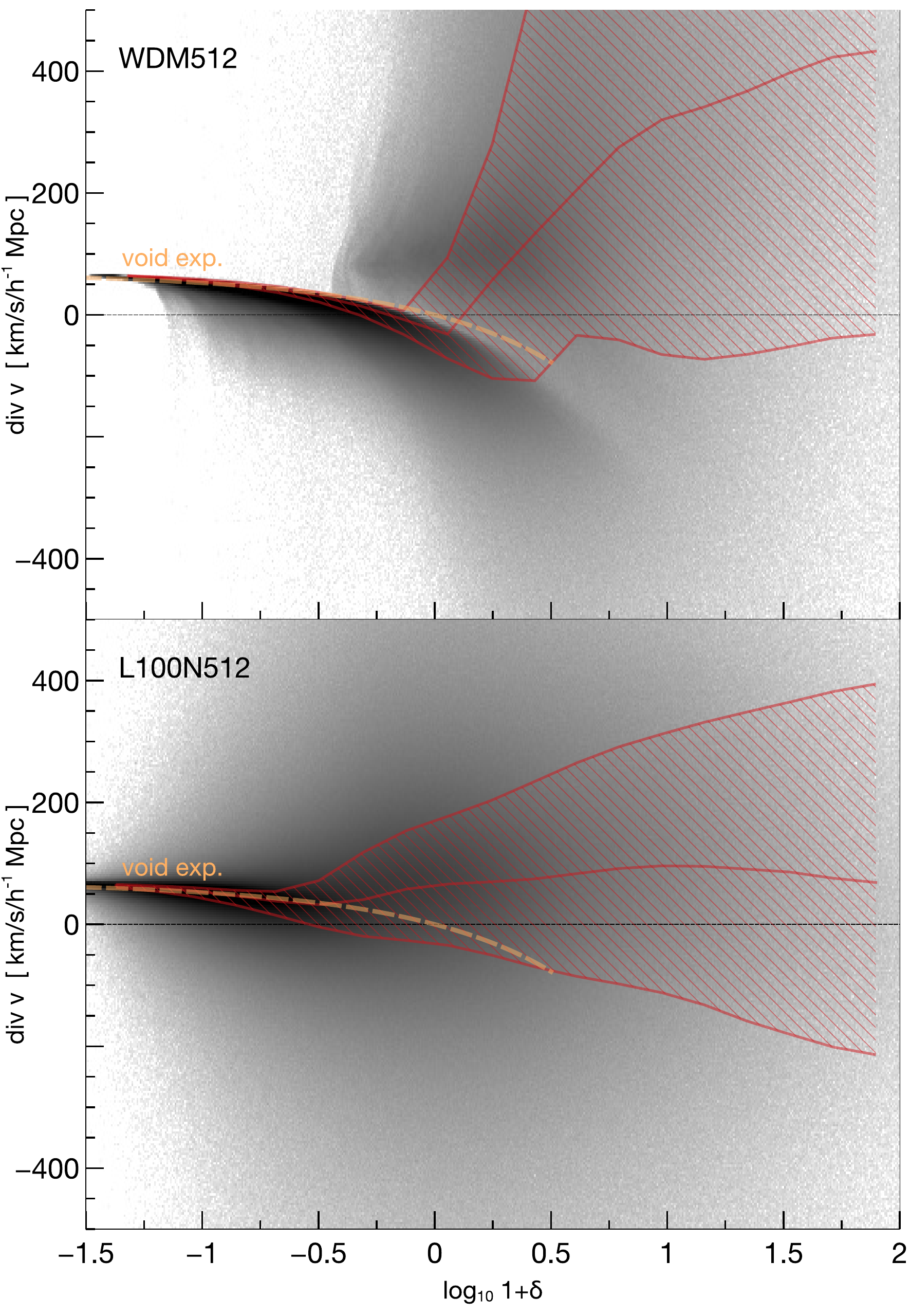}
\hspace{1.5cm}
\includegraphics[width=0.4\textwidth]{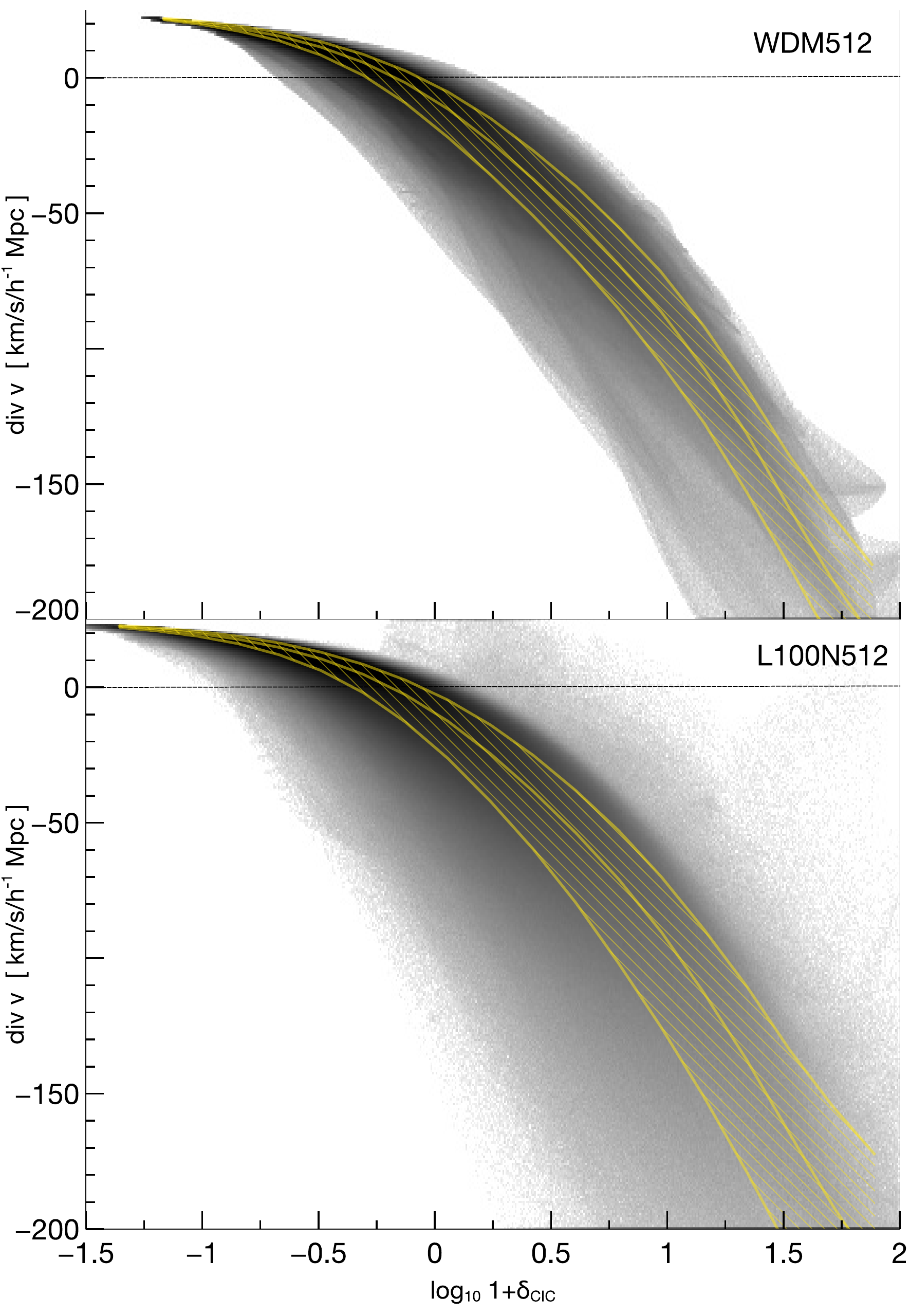}
\end{center}
\caption{\label{fig:hist_rho_divv}The correlation between velocity divergence and density.  {\bf Left Panel:} Here we show the point-wise velocity divergence and sheet density for the WDM512 simulation (top) and the L100N512 box (bottom). In the point-wise measurement, at low overdensities, a strong correlation exists that disappears around mean density. The red hatched area indicates the 16th and 84th percentile as well as the median in each bin of $\delta$. The dashed orange line represents the velocity divergence of a spherical void of overdensity $\delta$. {\bf Right Panel:} The spatially averaged version of the left panel, where the density and the velocity divergence field were smoothed with a Gaussian kernel on $1\,h^{-1}{\rm Mpc}$ scales for the WDM512 simulation (top) and the L100N512 box (bottom). We observe the well-known strong anti-correlation between density and velocity divergence. The yellow lines indicate median and 16th and 84th percentile. See text for details.
}
\end{figure*}

Most remarkably, {\em all} velocity divergence PDFs have a distinct feature, a
bump just below ${\rm div}\,\mathbf{v}\sim70\,{\rm km}/{\rm s}/h^{-1}\,{\rm
Mpc}$. The origin of this feature becomes apparent when we plot the two
dimensional distribution of overdensity-velocity~divergence pairs. We show
these, separately for the CDM simulation L100N512 and WDM512, in the left panel of
Fig.~\ref{fig:hist_rho_divv}. From this figure, and rather as expected, it is
readily apparent that in underdense regions, a tight correlation between
velocity divergence and overdensity exists. This correlation weakens with
increasing density, and completely disappears for
$\log_{10}\,1+\delta\gtrsim-0.5$ in the case of CDM, and for $\delta\gtrsim0$
for WDM. In the WDM case, we observe a very sharp upper limit to the possible
velocity divergence at a given underdensity. This limit can be easily
understood, since the expansion rate $\theta_{\rm void} \equiv H_0^{-1} {\rm
div}\, \mathbf{v}$ of a spherical void is directly related to its overdensity
$\delta$ (which, of course, is in fact an underdensity) through the nonlinear
equations \citep[cf.][]{Bernardeau:1997,Weygaert:2008} \begin{equation}
\theta_{\rm void} = \frac{3}{2}\frac{\Omega_m^{0.6} - \Omega_{\rm void}^{0.6}}{1+\Omega_{\rm void}^{0.6}/2};\quad \Omega_{\rm void} = \frac{\Omega_m\left(1+\delta\right)}{\left(1+\theta_{\rm void}/3\right)^2 }.
\end{equation}
We indicate the spherical void expansion rate by a dashed orange line in
Fig.~\ref{fig:hist_rho_divv}. It attains a maximum value of $\sim69\,{\rm
km}/{\rm s}/h^{-1}{\rm Mpc}$ for $\delta=-1$, which we show as the vertical
line in Figure~\ref{fig:velocity_pdfs}. In the CDM case, this does not in fact
provide an upper limit but rather describes the median relation for
$\delta\lesssim-0.5$ very well. This is plausibly simply due to the small-scale
noise inherent to CDM simulations.

The apparent loss of the correlation at mean density seems to contradict
the well known anti-correlation between density and velocity divergence that
is exact in linear theory and not expected to break down so dramatically at mean density.
The difference is, of course, that the exact formulae are point-wise and do {\em not} include
the contribution of caustics.  As discussed in Section~\ref{sec:derivatives}, the discontinuous
velocity jumps with associated singular derivative live on a subspace of measure zero and 
thus do not contribute to the volume-weighted one-point statistics without coarse-graining.
Once we smooth our density and velocity fields on some
scale so that the caustics are blurred out over a finite scale, we recover the expected result.
In the right panel panel of Fig.~\ref{fig:hist_rho_divv}, we thus show the
density against the velocity divergence computed using finite differences
from the sheet estimate of the mean velocity field. In contrast to the left panel, density and velocity
divergence were smoothed with a Gaussian filter on a $1\,h^{-1}{\rm Mpc}$ scale. This implies
that ordinary differentiation can be applied since the discontinuities disappear.
As expected, we recover the strong correlation that is expected between these
two quantities even in the non-linear regime \citep[cf. e.g.][]{Bernardeau:1999,Kudlicki:2000}.
Interestingly, the correlation in the WDM simulation is significantly tighter, but it is plausible that
the scatter is simply reduced due to the relatively small box size of that simulation.

Finally, we verify the global volume average of the divergence field.  We note that 
$\left<{\rm div}\,\mathbf{v}\right>=0$ (the angle brackets here denote a volume average)
in the linear regime -- due to Stokes' theorem 
and periodic boundary conditions -- and it is interesting to ask whether this holds
in the non-linear regime as well, despite the markedly asymmetric PDFs that we 
found above. First, for the divergence of the smoothed field this is true by definition,
since in Fourier space the derivative operator is zero at $\mathbf{k}=0$, implying
a vanishing DC mode. Second, for the exact divergence estimated on the sheet, we can
only compute this numerically. We find that for the WDM runs the volume averaged
mean divergence is about 0.5 per cent of the standard deviation of the divergence, while
for the highest resolution CDM run, i.e. L100N512, it is $\sim1.5$ per cent and 
much smaller for the larger CDM boxes, i.e. overall consistent with a vanishing mean.

We thus arrive at a rather interesting conclusion. The big difference between the density--velocity divergence
relation including jumps at caustics and excluding them demonstrates quite clearly that caustics
with their compressive jumps dominate the velocity statistics in overdense regions when coarse
graining is applied. On the other hand, if the compressive caustics are not included, the velocity field
in overdense regions is predominantly of positive divergence, completely consistent with our discussion
of the velocity field of the single halo in Section~\ref{sec:single_halo}. Another interesting observation is that the maximum void expansion rate
is not reached asymptotically in the smoothed statistics.


\section{Spectral properties of the cosmic velocity field}
\label{sec:kspaceresults}
In this section, we present our main results regarding the {\em spectral} properties of cosmic velocity fields. We first outline fundamental differences between divergence and vorticity calculated in real and in Fourier space, before we will present power spectra of the velocity divergence and vorticity as well as density-divergence cross spectra for all our CDM and WDM simulations along with a convergence study and a comparison with results obtained with the DTFE. 


\subsection{Radial and transversal velocity modes vs. divergence and curl}
\label{sec:velocity_modes}
The Fourier transform of the gradient operator $\boldsymbol{\nabla}$ is given by $-i\mathbf{k}$, so that in Fourier space the divergence and the curl of the velocity field $\mathbf{v}$ become purely radial and transversal projections of the Fourier transformed velocity field: 
\begin{eqnarray}
\tilde{\theta} &\equiv & -i\mathbf{k}\cdot\tilde{\mathbf{v}} \\
\tilde{\boldsymbol{\omega}} &\equiv& -i\mathbf{k}\times\tilde{\mathbf{v}},
\end{eqnarray}
where the tilde denotes a Fourier transformed field, which we will omit in what follows. 

We note that we do not use the divergence and curl computed using the sheet
to study the spectral properties, but rather the velocity field itself and then
perform the radial and transversal projections. The reason for this is that
a Fourier transform of the singular derivatives at the caustics needs to be
performed as well, rather than just the field away from the caustics in order 
to obtain the correct spectra. This result can be easily seen from the definition 
of the Fourier transform:
\begin{eqnarray}
\int_\mathbb{R} (v_r-v_l)\,\,\delta_D(x-x_o)\,\exp\left[-ikx\right]\,dx&& \nonumber\\
\qquad=\quad(v_r-v_l)\exp\left[-ikx_0\right].&&
\end{eqnarray}

One might be tempted to believe that only spectral derivatives should be
used and derivatives on the tetrahedra should be avoided. While this is
certainly warranted for the spectral properties of the velocity field, the
inverse discrete Fourier transforms of the fields $\theta$ and
$\boldsymbol{\omega}$ have little to do with the divergence and curl of the
mean velocity field. This is expected since the mean velocity fields
are discontinuous and thus slowly decaying in Fourier space so that their
Nyquist limited spectral derivatives are particularly ill behaved at the caustic
locations were the derivative is infinite.  We remind
the reader of the comparison of divergence fields obtained with a spectral
derivative, a 4th order accurate finite difference operator, as well as by
using 1st order accurate finite differences on the tetrahedra with explicit
evaluation of the derivative of the projected field (according to
eq.~\ref{eq:tet_divergence}) in Fig.~\ref{fig:divv_panels}.


\subsection{Power spectra of the radial and transversal modes and the cross-spectrum with density -- $P_{\theta\theta}$, $P_{\omega\omega}$ and $P_{\delta\theta}$}
\label{sec:velocity_properties}

In what follows, we are concerned with the power spectrum of the radial $\theta(\mathbf{k})$ and transversal velocity modes $\boldsymbol{\omega}(\mathbf{k})$ defined using spectral derivatives (see the discussion in Section~\ref{sec:velocity_modes}). Their respective power spectra are given by
\begin{eqnarray}
\left< \theta({\bf k})\, \theta^\ast({\bf k}')\right> &=& P_{\theta\theta}(k)\,\delta_D({\bf k} - {\bf k}'), \\
\left< {\boldsymbol{\omega}}({\bf k}) \cdot {\boldsymbol{\omega}}^\ast({\bf k}')\right> &=& P_{\omega\omega}(k)\,\delta_D({\bf k} - {\bf k}'). 
\end{eqnarray}

Cosmological initial conditions generated at arbitrary order of Lagrangian perturbation theory represent a purely potential flow before shell-crossing\citep[see e.g.][for a review]{Bernardeau:2002}, i.e. $\mathbf{v} = \boldsymbol{\nabla}\psi$, where $\psi$ is the velocity potential at a given order of the perturbation theory. Hence, a Poisson equation exists for $\psi$ of the form $\Delta\,\psi = 4\pi \theta$ and at linear order $\theta\propto-\delta$. However, at late times, vorticity is generated in the mean velocity field by multi-streaming, just as we have discussed in the first part of this paper.

It is thus interesting to consider the cross-spectrum between the overdensity $\delta$ and the velocity divergence $\theta$, i.e.
\begin{equation}
\left< \theta({\bf k})\, \delta^\ast({\bf k}')\right> = P_{\theta\delta}(k)\,\delta({\bf k} - {\bf k}'),
\end{equation}
\noindent In linear perturbation theory, this is just a scaled version of the density
power spectrum, but the growth of vorticity destroys the potential flow and
transversal modes become important. We thus expect the cross-spectrum to drop
rapidly at the scales where vorticity is generated and to reverse its sign on
scales where shell-crossing has occurred, which is exactly what we will find
below.

As we have discussed in AHK12 and \cite{Hahn:2012}, the density field estimated
from the sheet is somewhat biased high in the densest regions (inner regions of
haloes) since the sheet, linearly interpolated between particles, no longer
tracks there the true distribution function. As we want to complement the
velocity field analysis in this paper with an unbiased estimator, we resort to
the noisy standard method of simple CIC deposit into the three-dimensional data
cube \citep{Hockney:1981}. We then deconvolve with the CIC assignment kernel to
correct the suppression of small-scale power close to the grid Nyquist wave
number. Specifically, we deconvolve with the kernel
\begin{equation*}
W_{\rm CIC}(\mathbf{k}) = \prod_{i=x,y,z}\left(\frac{\sin \pi k_i/2k_{\rm Ny}}{\pi k_i/2k_{\rm Ny}}\right)^2
\end{equation*}
before computing the  power spectrum of the overdensity field $\delta$ by evaluating
\begin{equation}
\left< \delta({\bf k})\, \delta^\ast({\bf k}')\right> = P_{\delta\delta}(k)\,\delta_D({\bf k} - {\bf k}').
\end{equation}

We compute all spectra by performing a fast Fourier transformation of the three-dimensional data cube 
of the respective quantities followed by computing binned averages of all modes that fall into
 logarithmically spaced intervals $k_j\leq k\leq k_{j+1}$.


\subsection{Velocity divergence and vorticity power spectra for WDM}

\begin{figure}
\begin{center}
\includegraphics[width=0.42\textwidth]{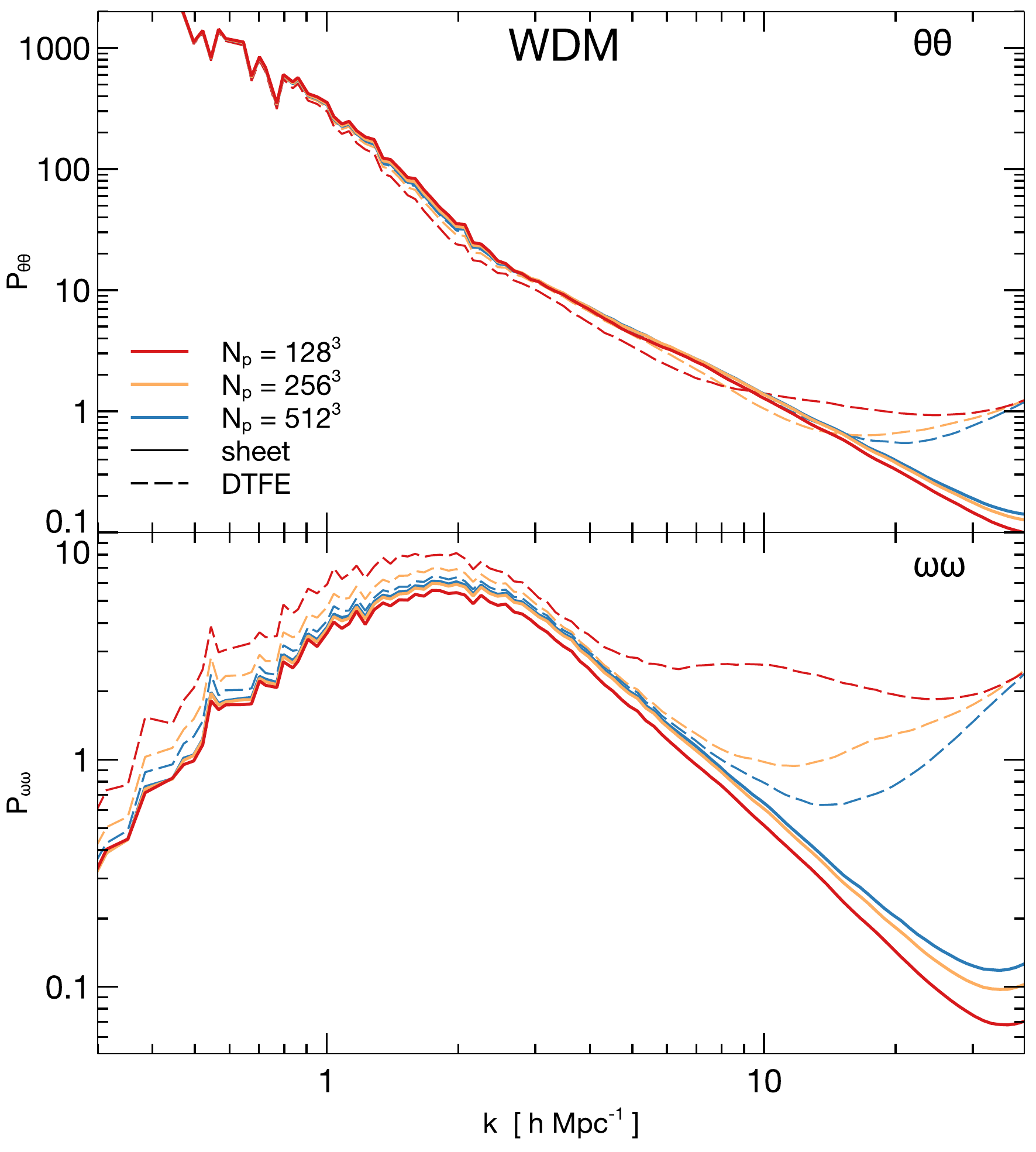}
\end{center}
\caption{\label{fig:spectra_wdm}Power spectra at $z=0$ of the divergence (top
panel) and the vorticity (bottom panel) for the WDM simulations. All
simulations used the same PM force resolution of $512^3$ cells, but had varying
mass resolution: $128^3$ (red), $256^3$ (yellow) and $512^3$ (blue). Results
obtained using the sheet tessellation are shown with a {\em solid line},
respective results obtained with {\sc dtfe} are shown with {\em dashed lines}.
While the two methods yield comparable results on large scales, only the sheet
method shows convergence on small scales.  }
\end{figure}

We first investigate the velocity divergence and vorticity power spectra for
our WDM simulations, which all have the same PM force resolution but have
varying mass resolution. The $300\,{\rm eV}$ initial WDM power spectrum leads
to a truncation of perturbations that is resolved in all simulations \citep[see
also the discussion in][]{Angulo:2013}. At fixed force resolution this should
mean that one can arrive at perfectly converged properties. This is in stark
contrast to the CDM case that we will discuss below, where more perturbations
are introduced when the mass resolution is increased since the perturbation
spectrum of CDM continues to almost arbitrarily small scales (see also our
discussion of this in AHK12 and \citealt{Hahn:2012})

For this reason, we expect to achieve actual convergence if the simulation
resolves both the cut-off scale and the non-linear scale. We thus re-analyse
the simulations introduced in \cite{Hahn:2012} for a toy model $300\,{\rm eV}$
WDM particle. We do not use the simulation outputs obtained with the new
simulation technique introduced in that paper, but make use of the standard PM
results in order to keep the discussion purely focused on the analysis of the
simulation data rather than the simulation technique. 
 
We show the divergence and vorticity power spectra for our three WDM runs in
Fig.~\ref{fig:spectra_wdm}, comparing again with {\sc dtfe}. It is quite
remarkable that the divergence spectra determined using the sheet are converged
irrespective of resolution at all $k\lesssim 10 \,h/{\rm Mpc}$ with very small
discrepancy at larger $k$. The vorticity spectra show a small resolution
dependence in amplitude, increasingly so at larger $k$, but not in shape. For
the divergence spectra obtained with {\sc dtfe}, we observe slower convergence,
in particular scales $k\gtrsim 3 \,h/{\rm Mpc}$ show a very strong dependence
on the mass resolution. Even more pronounced is the lack of convergence for the
vorticity spectra. Here, in the case of {\sc dtfe}, all scales show a
resolution dependence, unlike for the sheet estimated vorticity which is
perfectly converged on small scales. In addition, for {\sc dtfe}, at small
scales, the vorticity spectra appear to converge slowly to a power-law with
positive index. This is most likely due to the lack of an actual projection of
the distribution function in this approach, where the single particle velocity
dispersion becomes sampled in multi-stream regions rather than the mean
velocity. These results are a particularly good example of the strength of our
phase-space based estimate of cosmic velocity fields. A remarkable difference
is that the phase space sheet estimate of the vorticity spectrum converges from
below, while the {\sc dtfe} estimate converges from above. 

\begin{figure}
\begin{center}
\includegraphics[width=0.4\textwidth]{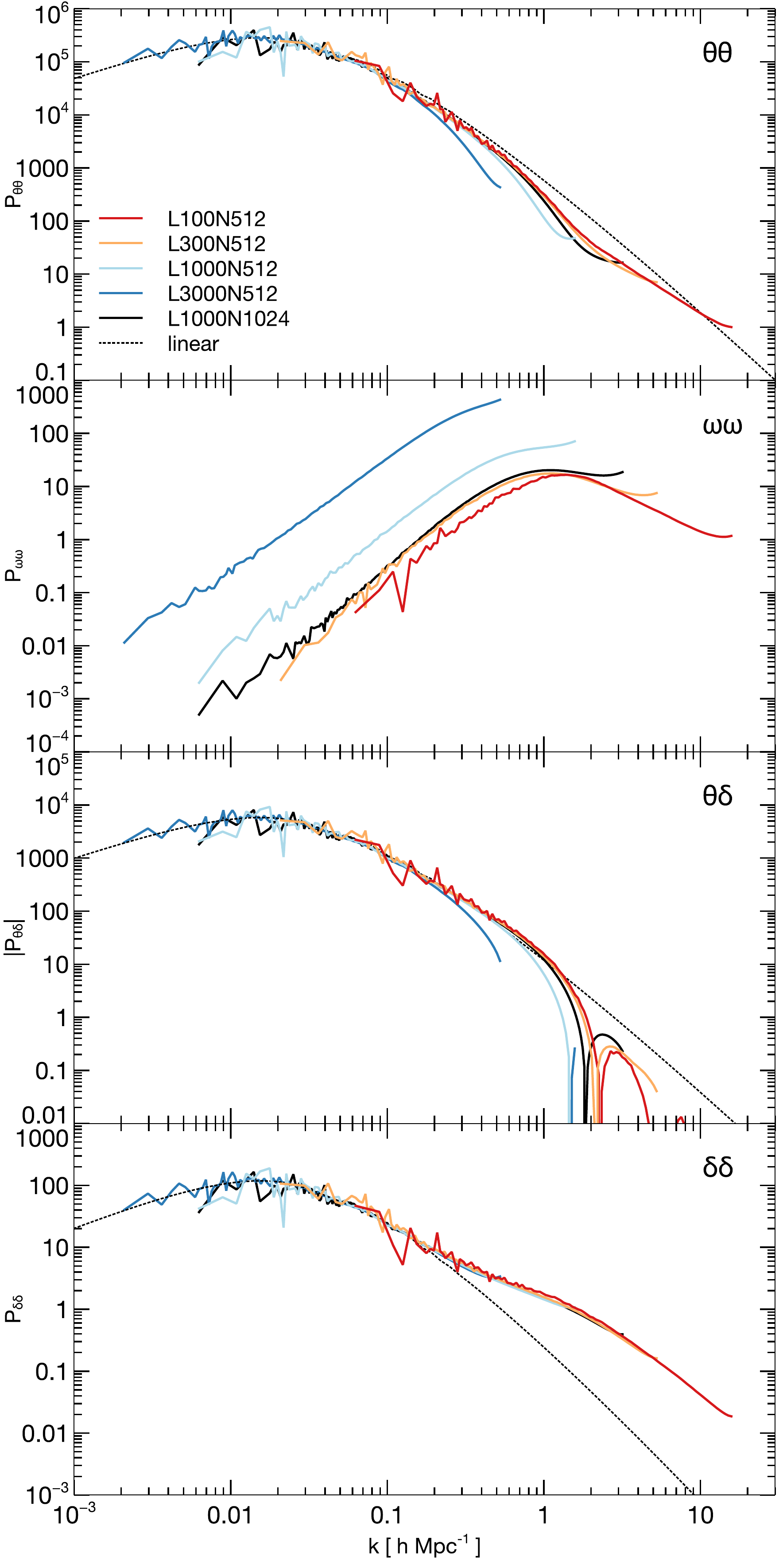}
\end{center}
\caption{\label{fig:velocity_spectra_boxes}Power spectra of the velocity divergence (top panel), vorticity (second from top), the cross spectrum between velocity divergence and overdensity (third from top) and the density power spectrum (bottom). Data is given for all the simulations we performed, indicated by the different colors. The spectra predicted by linear perturbation theory are given by the dotted lines, the prediction for the vorticity spectrum is zero.
}

\end{figure}
\begin{figure}
\begin{center}
\includegraphics[width=0.4\textwidth]{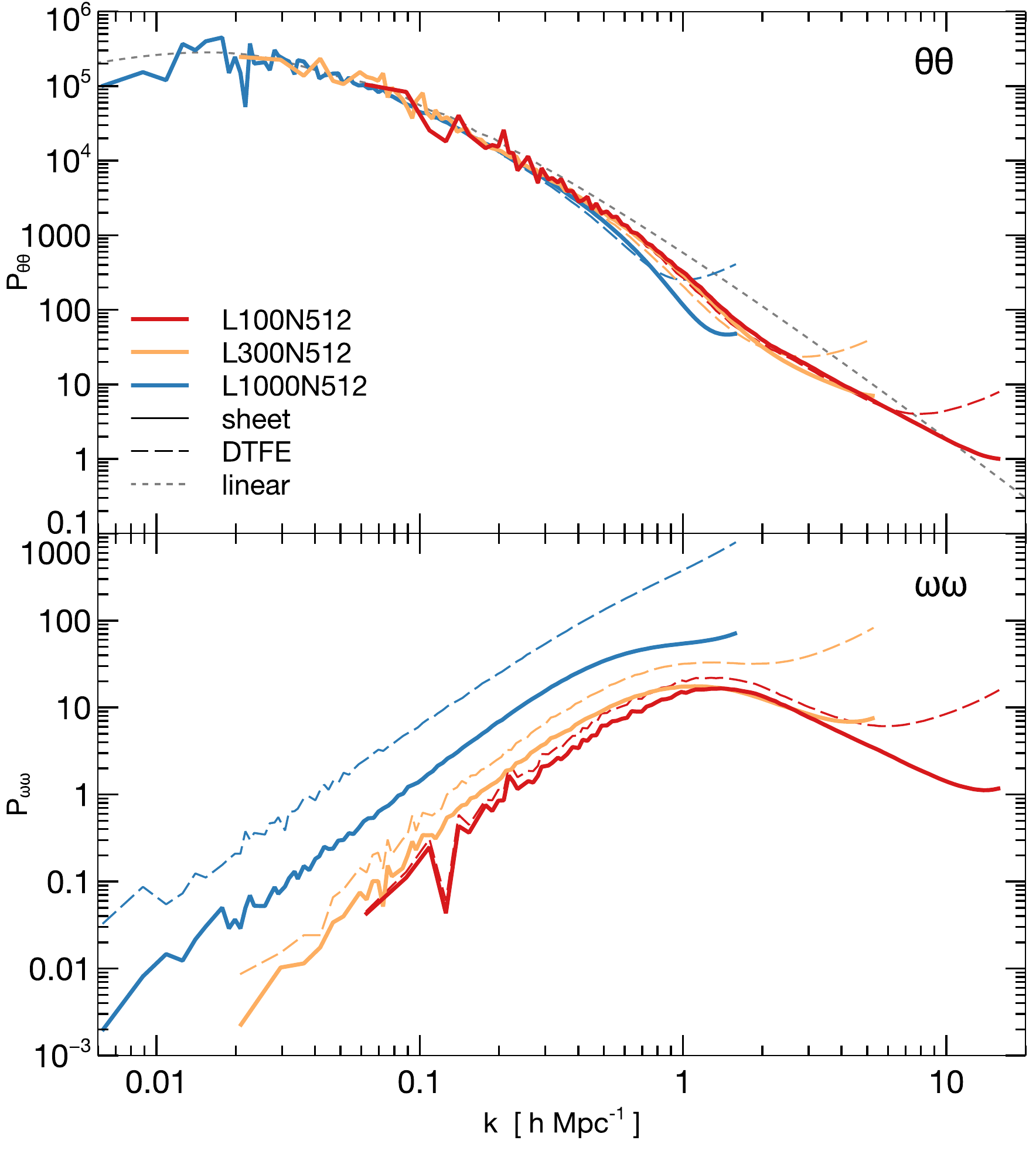}
\end{center}
\caption{\label{fig:spectra_dtfecomp}Power spectra at $z=0$ of the divergence (top panel) and the vorticity (bottom panel) of the $1\,h^{-1}{\rm Gpc}$ (red), $300\,h^{-1}{\rm Mpc}$ (yellow) and $100\,h^{-1}{\rm Mpc}$ (blue) boxes with $512^3$ particles in each case. Results obtained using the sheet tessellation are shown with a {\em solid line}, respective results obtained with {\sc dtfe} are shown with {\em dashed lines}.
}
\end{figure}

\subsection{Results: Power spectra of the velocity and density field for CDM}
In Fig.~\ref{fig:velocity_spectra_boxes}, we present, for all 5 CDM
simulations, the velocity divergence (top panel), and vorticity (second from
top panel) power spectra, as well as the density-velocity divergence cross
spectrum (third panel from top) and the density power spectrum (bottom panel).
Our results are largely consistent with results of \cite{Pueblas:2009}. In
particular, we observe as well that converged power spectra require that the
non-linear scale is well resolved. \cite{Pueblas:2009} observe that a mass
resolution below $10^9\,h^{-1}{\rm M}_\odot$ is necessary at $z=0$ for
converged spectra, consistent with our results. Both, the results of these
authors as well as ours indicate that this is mostly due to the slow
convergence of the flow vorticity. Somewhat unexpectedly however,
\cite{Pueblas:2009} do not find an obvious resolution dependence of the
convergence spectrum (while they do for the vorticity spectrum). 

Since vorticity is driven by small-scale non-linearities, we find that the slow
convergence of vorticity is, at least to some degree, driven by a lack of
resolution at the non-linear scale: when running the L1000N512 box with a force
resolution of only 8 times the mean inter-particle distance we observe a
significant drop of the large-scale vorticity spectrum (not shown).

We next discuss the differences we see in spectra estimated using the phase
space sheet compared to a standard Delaunay tessellation approach. In
Figure~\ref{fig:spectra_dtfecomp}, we compare the divergence and vorticity
spectra computed from the velocity field estimates based on the dark matter
sheet with those obtained using the {\sc dtfe}. We perform this comparison only
for the three simulations L100N512, L300N512 and L1000N512. For the divergence
spectrum, we observe that (1) all spectra, from {\sc dtfe} and sheet, converge
to the same spectrum with increasing small scale resolution, (2) the {\sc dtfe}
spectra show a pronounced rise at the smallest scales which is due to the small
scale noise discussed above in Section~\ref{sec:velocity_slices}, (3)
convergence at all scales is faster for the sheet-based estimates than for the
{\sc dtfe}-based estimates. For the vorticity spectra, which are clearly the
most challenging to estimate reliably,  an essentially identical picture
emerges but convergence is considerably slower with increasing resolution for
{\sc dtfe} than for the sheet based estimate.

\begin{figure}
\begin{center}
\includegraphics[width=0.4\textwidth]{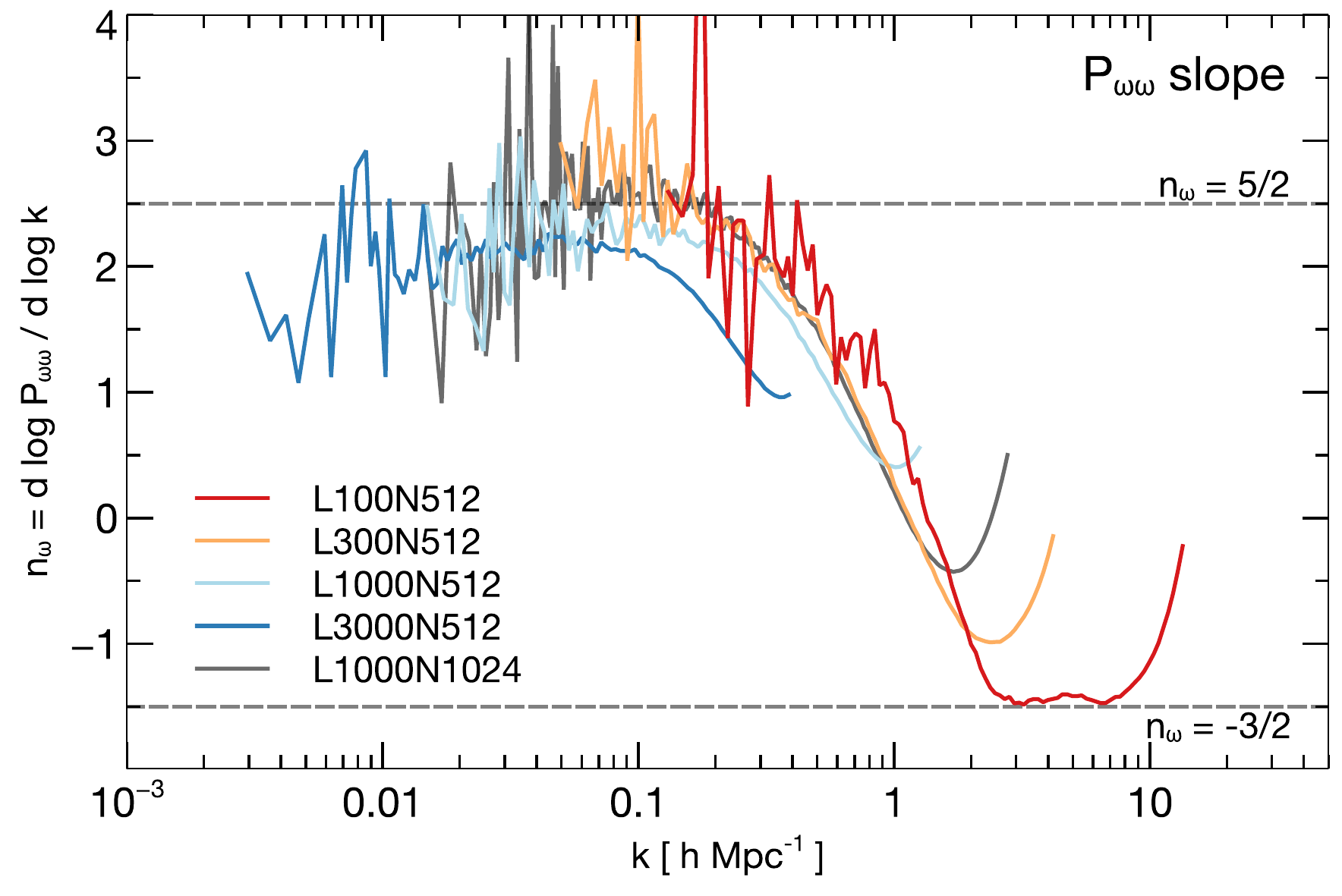}
\end{center}
\caption{\label{fig:vorticity_slope}Slope of the vorticity power spectrum $P_{\omega\omega}$ at $z=0$ showing an asymptotic behaviour consistent with $P_{\omega\omega}\propto k^{5/2}$ for small $k$ and $P_{\omega\omega}\propto k^{-3/2}$ for large $k$.
}
\end{figure}

In Fig.~\ref{fig:vorticity_slope}, we show the slope of the vorticity power
spectrum $n_\omega\equiv{\rm d}\log P_{\omega\omega}/{\rm d}\log k$ as a
function of wavenumber $k$. We find that $n_\omega\to5/2$ on large scales for
$k\lesssim 0.1\,h\,{\rm Mpc}^{-1}$ (consistent with the predictions of
\citealt{Carrasco:2014}) and $n_\omega\to-3/2$ on small scales for
$k\gtrsim1\,h\,{\rm Mpc}^{-1}$. The simulations that have the lowest mass
resolution, notably L1000N512 and L3000N512, already showed a spurious
significantly enhanced vorticity amplitude, also have a slightly lower
$n_\omega$ on large scales. Unfortunately, only our simulation with the highest
mass resolution demonstrates the asymptotic behaviour of the vorticity on small
scales so that we cannot establish convergence of this value. This requires
further investigation with higher resolution simulations to robustly measure 
the small-scale asymptotic slope.


\subsection{Fitting formulae}

\begin{figure*}
\begin{center}
\includegraphics[width=0.4\textwidth]{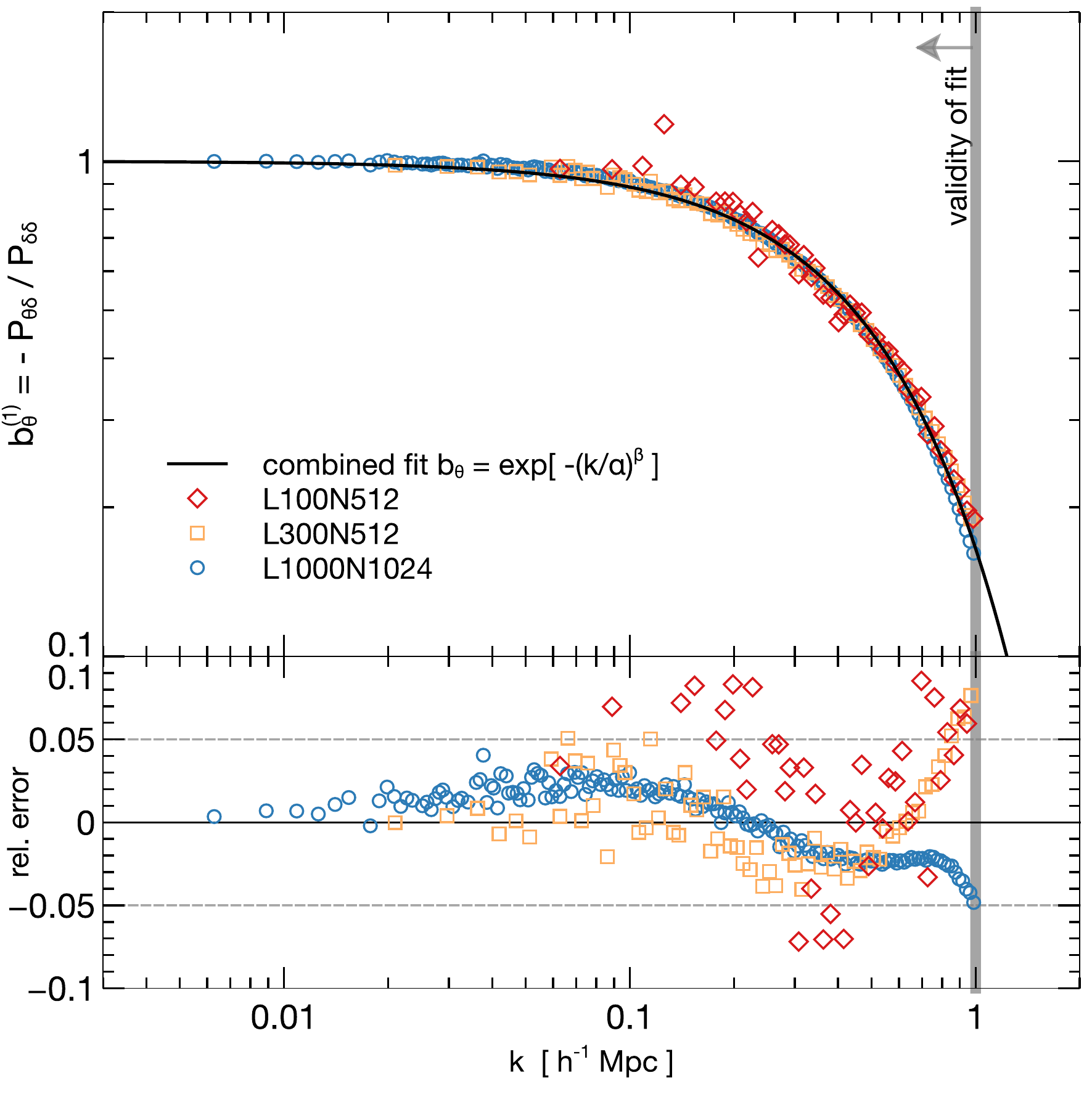}\qquad\qquad
\includegraphics[width=0.4\textwidth]{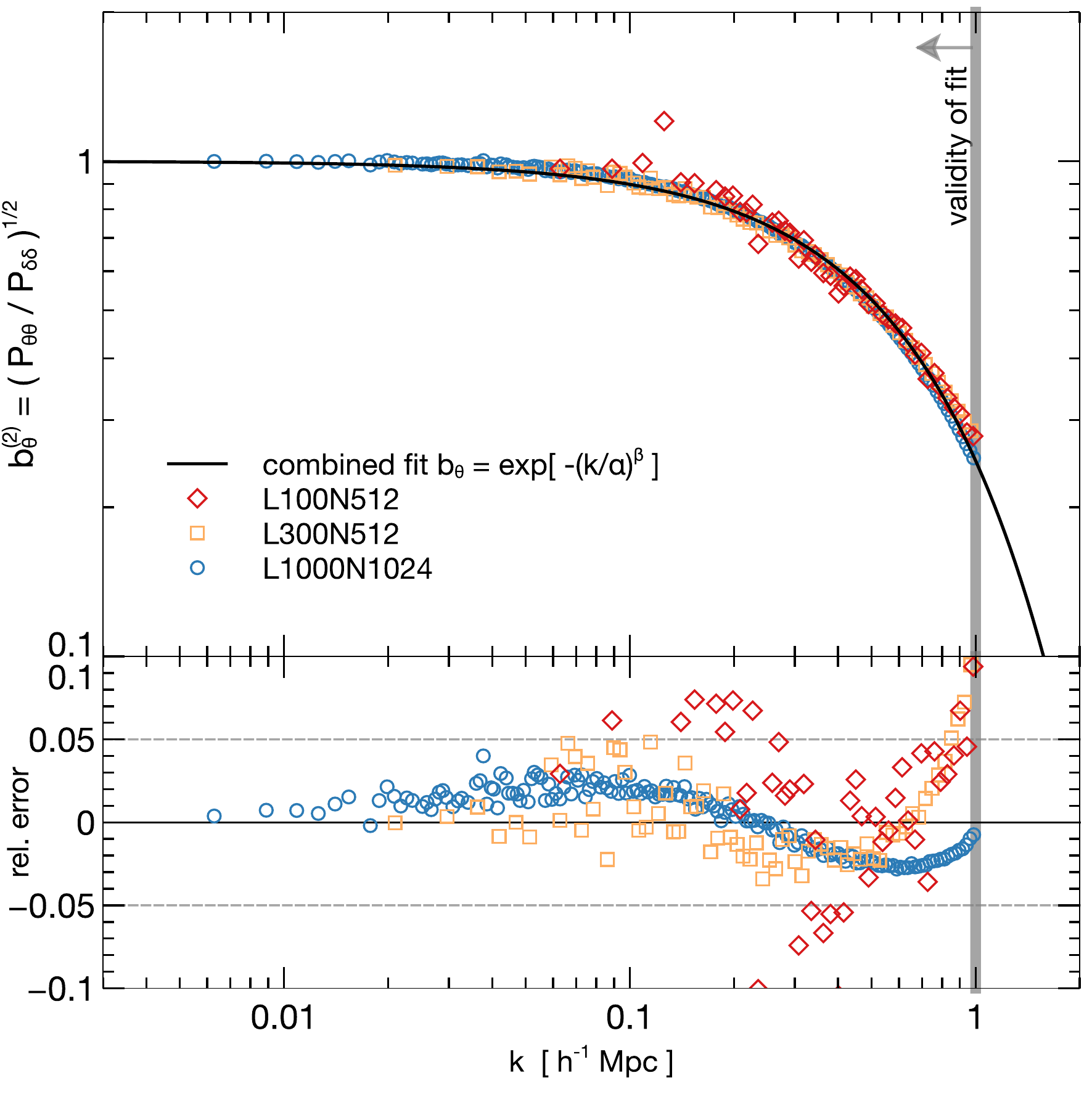}
\end{center}
\caption{\label{fig:bias_fit}Fit of the divergence bias $b_\theta$ from the
divergence-density cross spectrum (left panel) and from the divergence power
spectrum (right panel) at $z=0$ with an exponential function $b_\theta(k) =
\exp\left[ - (k/\alpha)^\beta\right]$ with best fitting parameters
$\alpha=0.606\pm0.004\,h\,{\rm Mpc}^{-1}$ and $\beta=1.176\pm0.017$ when
estimated from the cross spectrum (left), as well as
$\alpha=0.7423\pm0.006\,h\,{\rm Mpc}^{-1}$ and $\beta=1.112\pm0.018$ when
estimated from the divergence power spectrum (right). The vertical grey line
indicates $k=1\,h\,{\rm Mpc}^{-1}$, the wave number up to which our fit was
performed and is valid with an accuracy of about 5 per cent. At even larger
$k$, $b_\theta$ changes its sign (see Fig.~\ref{fig:velocity_spectra_boxes}).
}
\end{figure*}

In this section, we determine fitting functions for the bias of the velocity divergence relative to the matter density field for $k<1\,h\,{\rm Mpc}^{-1}$. This divergence bias $b_\theta$ can be defined either through the density--velocity divergence cross spectrum as
\begin{equation}
b_\theta^{(1)} = -\frac{1}{\mathcal{H}f}\frac{P_{\delta\theta}}{P_{\delta\delta}}
\label{eq:bias1}
\end{equation}
or through the velocity divergence power spectrum as
\begin{equation}
b_\theta^{(2)} = \frac{1}{\mathcal{H}f}\sqrt{\frac{P_{\theta\theta}}{P_{\delta\delta}}},
\label{eq:bias2}
\end{equation}
where $\mathcal{H}f \equiv {\rm d}\log D/{\rm d}\log \tau$ and $D(\tau)$ is the linear growth factor. We find that both are well fit for $k\lesssim1\,h\,{\rm Mpc}^{-1}$ by the exponential function
\begin{equation}
b_\theta(k) = \exp\left[ - \left(k/\alpha\right)^\beta \right].
\end{equation}
We determine the fit parameters $\alpha$ and $\beta$ by combining the spectra
of the three boxes L100N512, L300N512 and L1000N512 for $k<1\,h\,{\rm
Mpc}^{-1}$. The divergence bias calculated from eq.~(\ref{eq:bias1}) is shown
in Fig.~\ref{fig:bias_fit} (top panel, left) for the three boxes, the one
calculated from eq.~(\ref{eq:bias2}) in the top panel, right. When fitting the
bias from the cross-spectrum, we find best fitting parameters
$\alpha=0.606\pm0.004\,h\,{\rm Mpc}^{-1}$ and $\beta=1.176\pm0.017$. When
fitting from the power spectrum, the best fit parameters are
$\alpha=0.7423\pm0.006\,h\,{\rm Mpc}^{-1}$ and $\beta=1.112\pm0.018$. The
relative error with respect to the fits is shown in the bottom panels of
Fig.~\ref{fig:bias_fit}. We find that for $k\lesssim1\,h\,{\rm Mpc}^{-1}$, the
fits describe the data with about 5 per cent accuracy or better on larger
scales. For a linear bias in the absence of noise, the two bias determinations
should yield the same answer, but we find that this is only approximately true.
Particularly, the parameter $\alpha$ implies a cut-off at slightly larger
scales for the bias calculated from the cross-spectrum. We note that the
density fields, obtained with CIC deposits, contain a shot-noise term that we
have not subtracted. The level of agreement between the different simulations
(with different amplitudes of shot-noise) implies however that the discrepancy
between the two biases cannot be explained by shot-noise alone. The difference
between the fits is thus plausibly caused by either a non-linear relation
between density and velocity-divergence or another more complicated additive
noise component.
The difference between the fits could in part be due to a more complicated additive
noise component, but most likely reflects the different scale-dependence of the non-linearities
in density and velocity divergence \citep[see][who find that $b_1(k)/b_2(k)$ is monotonically decreasing]{Ciecielg:2004}. 

\section{Summary and Conclusions}
\label{sec:summary}

In this paper, we have extended the phase-space sheet tessellation method
introduced in AHK12 to estimate the properties of volume-weighted cosmic
velocity fields and their differentials, explicitly, the velocity divergence
and vorticity. To our knowledge, this method is the only one that 
is able to take into account the discontinuous and multi-valued 
nature of the velocity field in multi-stream regions that arise wherever
structures collapse gravitationally.

Even {\sc dtfe}, which so far has been perhaps 
the best velocity estimator for numerical simulations, is plagued by 
noise in multi-stream regions and shows convergence relatively slowly 
for all properties of the velocity field. This can naturally be 
circumvented by our method, without the need of filtering on relatively 
large-scales at the expense of loosing detailed features.

\vspace{0.2cm}
\noindent We can summarise the advantages of our approach as follows:
\begin{enumerate}[(i)]
\item The phase space sheet allows a proper definition of a projection operator
from phase space onto configuration space and thus properly averaged velocity
fields in multi-stream regions. An estimate that does not respect the
phase-space connectivity leads to small-scale jitter by interpolating between
velocities that are close in configuration space but not close in phase space.
Using the correct phase-space connectivity significantly reduces small-scale
noise in the velocity fields.  
\item We derived exact expressions for the divergence and curl of the mean 
velocity field in multi-stream regions, eqs.~(\ref{eq:tet_divergence}) and
(\ref{eq:tet_curl}), that explicitly do not include the discontinuities of the
velocity field at caustics in multi-stream regions where derivatives are singular. 
\item By excluding the singular derivatives (which occupy a volume of measure zero), 
our technique can compute the differential properties of the mean velocity field without
coarse-graining and thus the introduction of an arbitrary scale. By doing so, the
differentials represent only the mean dynamics of the dark matter fluid instead of 
including also the motion of the singular caustics.
\end{enumerate}
Thus, in summary, we find that our new method provides significantly 
improved estimates of cosmic velocity fields and their differential properties: 
it allows a proper definition of a phase-space projection operator that 
significantly reduces small-scale noise. 

\vspace{0.2cm}
\noindent 
We applied this estimator to a set of cold and warm dark matter N-body simulations. We 
discussed in detail the differences that arise when ordinary finite 
differencing schemes and spectral derivatives are used. We showed 
that those operators perform an implicit coarse graining that includes the
singular derivatives. We also presented an in-depth discussion
of the real-space properties of the velocity field and its differential 
properties both in CDM and WDM. In the final part of the paper we
focused on the spectral properties. 

\vspace{0.2cm}
\noindent Our science results can be summarised as:
\begin{enumerate}[(i)]
\item We demonstrated explicitly that the mapping between density and velocity divergence, i.e. ${\rm div}\,\mathbf{v}\propto -\delta$, which holds in linear perturbation theory, appears as a strong correlation before shell crossing. After shell-crossing, the correlation reverses its sign. In shell-crossed regions, the only locations of predominantly negative divergence are the very centres of filaments and haloes. This implies that in overdense regions, the coarse-grained non-linear density--velocity anti-correlation is predominantly driven by the compressive caustics.
\item Our results from WDM simulations indicate that, inside of haloes, the spherically averaged vorticity drops sharply and the velocity divergence changes its sign around the location where the density profile has a slope of $-2$. The outer regions of the halo have a positive divergence, while the inner regions are convergent. This tentative result certainly warrants further investigation.
\item We discussed the 1-point statistics of the divergence and vorticity fields and found that the divergence PDF exhibits a pronounced feature at the maximum void expansion rate. The vorticity PDF is reasonably well fit by a lognormal distribution. Furthermore, the two-dimensional histogram of overdensity vs. velocity divergence shows that for CDM the correlation between overdensity and velocity divergence is well described by the void expansion rate for $\delta\lesssim-0.7$ while for WDM, the void expansion rate provides a sharp upper limit to the velocity divergence at a given overdensity for $\delta\lesssim-0.7$.
\item Velocity spectra for CDM obtained with the phase space sheet show faster convergence behaviour with resolution, but both our method and {\sc dtfe} converge to the same spectra when the non-linear scale is very well resolved. Consistent with previous results, we find that the vorticity spectra require higher resolution in order to be converged than divergence spectra.
\item We complemented our discussion of spectral properties with an analysis of WDM simulations, where the entire perturbation spectrum can be resolved. Here, we are able to estimate spectra that are almost independent of the resolution of the underlying simulation, while the estimate on the position-space tessellation shows only slow convergence with resolution, is dominated by small scale noise and lacks convergence to the correct spectrum at small scales.
\item Finally, we provided fits for a bias parameter $b_\theta$ in terms of a function exponentially decaying in wave number $k$ that allows the CDM divergence power spectrum $P_{\theta\theta}$ as well as the divergence--overdensity cross spectrum $P_{\theta\delta}$ to be related to the non-linear matter power spectrum $P_{\delta\delta}$.
\end{enumerate}

\noindent It will be interesting to follow up this preliminary analysis by probing the effect of cosmic large-scale velocity fields on the assembly of haloes and thus the alignment of their spins with larger scales as well as the connection between velocity fields and gravitational tidal fields\citep[c.f.][]{Aragon-Calvo:2007a,HahnPorciani:2007,Hahn:2007,Hoffman:2012,Libeskind:2013,Laigle:2015}. Another application of our method is to more reliably measure the ``effective field'' properties of dark matter \citep[e.g.][]{Carrasco:2012}. Future work should also study the second moment of the projected velocity field, i.e. the  anisotropic stress tensor, in addition to the properties of the mean field.


\section*{Acknowledgements}
O.H. thanks Aseem Paranjape for many fruitful discussions. The authors thank him as well as 
St\'ephane Colombi and Christophe Pichon for valuable
comments on the draft of this article. O.H. acknowledges support from the Swiss National Science
Foundation (SNSF) through the Ambizione fellowship. We are grateful to Rien van
de Weygaert and Marius Cautun for making the {\sc dtfe} software publicly
available. We gratefully acknowledge the support of Stuart Marshall and the
SLAC computational team, as well as usage of the computational resources at
SLAC and ETH Zurich. We thank our referee Michal Chodorowski for his
very valuable comments that helped to improve the presentation of this paper.



\appendix

\section{Derivation of the velocity divergence and curl in multi-stream regions}
\label{sec:derivation_divrot}
We briefly derive here the expressions ${\rm div }\left< \mathbf{v}\right>$ as given in eq.~(\ref{eq:tet_divergence}) 
and  ${\rm curl }\left< \mathbf{v}\right>$ as given in eq.~(\ref{eq:tet_curl}). This is achieved by applying the derivative
operator to the multi-stream average as given on the right-hand-side of eq.~(\ref{eq:vel_average}). Unlike in the 
main text, we write the stream index $s$ as a superscript in order to distringuish it more clearly from Cartesian indices
which are written as subscripts. We first note
that in order for the derivative to commute with the sum over streams, $\sum_s$, the number of streams is not allowed to change
in the point where the derivative is taken. Then $\partial_i\sum_s x^s = \sum_s \partial_i x^s$. 
This is only the case at caustics which is the reason for excluding them from
the derivatives and consider them separately in Section~\ref{sec:derivatives}. So, away from caustics, we now evaluate
the mixed derivative $\partial_i \left< v_j \right>$ from which divergence and curl can be constructed.We then have
\begin{eqnarray}
\partial_i \left< v_j \right> & = & \partial_i \frac{\sum_s \rho^s v_j^s}{\sum_s \rho^s} \nonumber \\
& = & \frac{\sum_s \left( v_j^s \partial_i \rho^s +\rho^s \partial_i v_j^s \right)}{\sum_s\rho^s}  - \frac{\left(\sum_s\rho^s v_j^s\right)\left(\sum_s\partial_i\rho^s\right)}{\left(\sum_s\rho^s\right)^2} \nonumber\\
& = & \frac{\sum_s v_j^s\rho^s\,\partial_i\log\rho^s}{\sum_s\rho^s} + \left< \partial_i v_j \right> -\left< v_j \right> \,\left< \partial_i \log \rho \right>\nonumber\\
& = & \left< \left(\partial_i \log \rho \right)\left(v_j - \left< v_j\right>\right)\right> + \left< \partial_i v_j\right>,
\label{eq:derivative_tensor}
\end{eqnarray}
where we have exploited various times the definition of stream averages $\left< x \right> = \left(\sum_s x^s \rho^s\right)/\sum_s \rho^s$ and the idempotence of this average $\left<\left<x\right>\right>=\left<x\right>$. The expressions for divergence and curl follow straightforwardly from the tensorial derivative (\ref{eq:derivative_tensor}).


\section{Resolution study of the velocity field in haloes}
\label{sec:resolution}
\begin{figure}
\begin{center}
\includegraphics[width=0.4\textwidth]{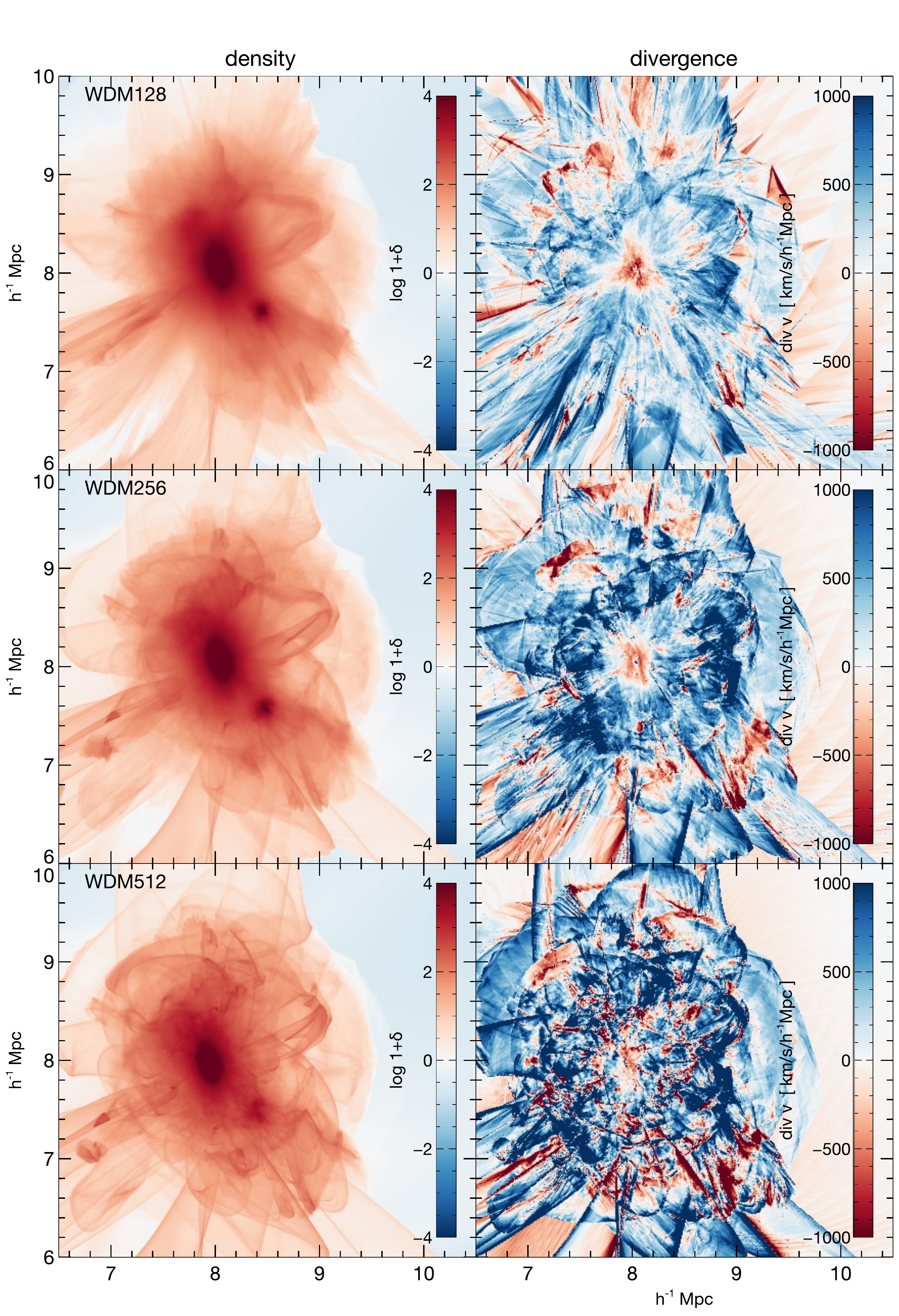}
\end{center}
\caption{\label{fig:halo_res_test}Convergence study of slices of the
dark-matter sheet estimated density field and velocity divergence through the
centre of the most massive halo of the WDM simulations. The bottom row panels
are a close-up version of the corresponding panels in
Fig.~\ref{fig:halo_slices}.  }
\end{figure}

In this appendix, we assess the degree of convergence of the velocity fields
estimated from the dark-matter sheet inside of virialized structures. We show a
close-up version of the most massive halo in the WDM simulations in
Fig.~\ref{fig:halo_res_test}, to be compared to the respective panels of
Fig.~\ref{fig:halo_slices}. We observe that the larger scale structures in the
density field are remarkably stable with increasing resolution. Convergence in
the velocity divergence field is however considerably slower.

\begin{figure}
\begin{center}
\includegraphics[width=0.4\textwidth]{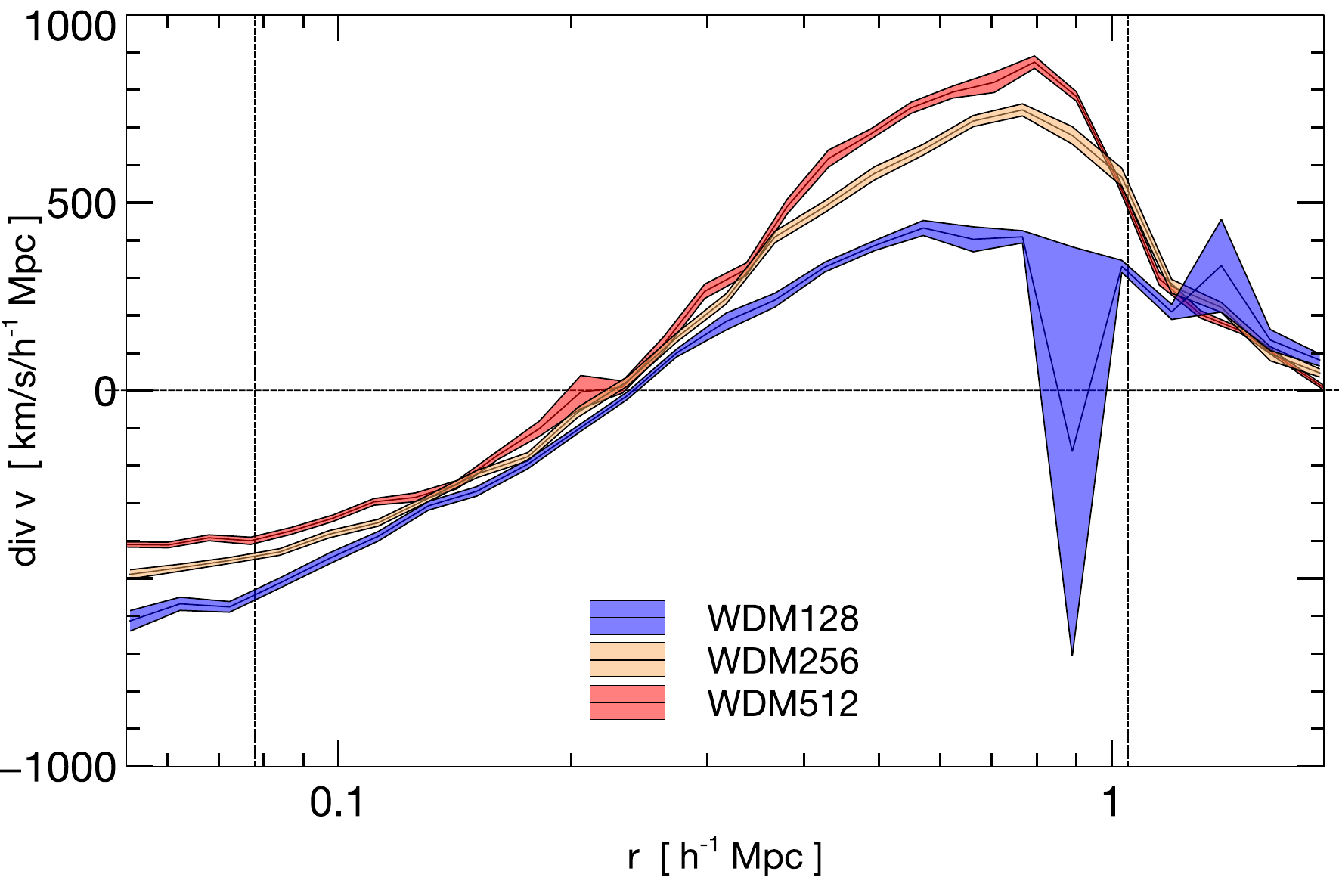}
\end{center}
\caption{\label{fig:profile_res_test}Convergence study of the spherically averaged single stream velocity divergence. Profiles are shown for all WDM simulations. While the peak properties have not yet converged, the point where the velocity field changes from convergent to divergent flow is consistent among all resolutions. Shaded regions correspond to the error on the median estimated from the 16th and 84th percentile in each radial bin.
}
\end{figure}

We complement this rather qualitative comparison with radial 
velocity divergence profiles of the averaged single-stream velocity
divergence in Fig.~\ref{fig:profile_res_test}. Consistent with the slices
discussed above, the qualitative features are stable across resolutions. While
the velocity divergence profile is not yet converged, the peak location as well
as the location of sign flip are consistent among the different resolutions.

\label{lastpage}

\end{document}

%% file: local-commands.tex
\def\k#1 {k_{{\rm #1}}}

\def\mH2p{H_2^+}

\def\ltsima{$\; \buildrel < \over \sim \;$}
\def\simlt{\lower.5ex\hbox{\ltsima}}   
\def\gtsima{$\; \buildrel > \over \sim \;$}
\def\gtsim{\lower.5ex\hbox{\gtsima}}

%% file: paper.bbl
\begin{thebibliography}{}

\bibitem[\protect\citeauthoryear{{Abel}, {Hahn} \& {Kaehler}}{{Abel}
  et~al.}{2012}]{Abel:2012}
{Abel} T.,  {Hahn} O.,    {Kaehler} R.,  2012, {\bf AHK12}, \mnras, 427, 61

\bibitem[\protect\citeauthoryear{{Angulo}, {Hahn} \& {Abel}}{{Angulo}
  et~al.}{2013}]{Angulo:2013}
{Angulo} R.~E.,  {Hahn} O.,    {Abel} T.,  2013, \mnras, 434, 3337

\bibitem[\protect\citeauthoryear{{Angulo}, {Springel}, {White}, {Cole},
  {Jenkins}, {Baugh} \& {Frenk}}{{Angulo} et~al.}{2012}]{Angulo:2012}
{Angulo} R.~E.,  {Springel} V.,  {White} S.~D.~M.,  {Cole} S.,  {Jenkins} A.,
  {Baugh} C.~M.,    {Frenk} C.~S.,  2012, \mnras, 425, 2722

\bibitem[\protect\citeauthoryear{{Angulo}, {Springel}, {White}, {Jenkins},
  {Baugh} \& {Frenk}}{{Angulo} et~al.}{2012}]{Angulo:2012a}
{Angulo} R.~E.,  {Springel} V.,  {White} S.~D.~M.,  {Jenkins} A.,  {Baugh}
  C.~M.,    {Frenk} C.~S.,  2012, \mnras, 426, 2046

\bibitem[\protect\citeauthoryear{{Arag{\'o}n-Calvo}, {van de Weygaert}, {Jones}
  \& {van der Hulst}}{{Arag{\'o}n-Calvo} et~al.}{2007}]{Aragon-Calvo:2007a}
{Arag{\'o}n-Calvo} M.~A.,  {van de Weygaert} R.,  {Jones} B.~J.~T.,    {van der
  Hulst} J.~M.,  2007, \apjl, 655, L5

\bibitem[\protect\citeauthoryear{{Ascasibar}}{{Ascasibar}}{2010}]{2010CoPhC.181.1438A}
{Ascasibar} Y.,  2010, Computer Physics Communications, 181, 1438

\bibitem[\protect\citeauthoryear{{Bernardeau}, {Chodorowski}, {{\L}okas},
  {Stompor} \& {Kudlicki}}{{Bernardeau} et~al.}{1999}]{Bernardeau:1999}
{Bernardeau} F.,  {Chodorowski} M.~J.,  {{\L}okas} E.~L.,  {Stompor} R.,
  {Kudlicki} A.,  1999, \mnras, 309, 543

\bibitem[\protect\citeauthoryear{{Bernardeau}, {Colombi}, {Gazta{\~n}aga} \&
  {Scoccimarro}}{{Bernardeau} et~al.}{2002}]{Bernardeau:2002}
{Bernardeau} F.,  {Colombi} S.,  {Gazta{\~n}aga} E.,    {Scoccimarro} R.,
  2002, \physrep, 367, 1

\bibitem[\protect\citeauthoryear{{Bernardeau} \& {van de
  Weygaert}}{{Bernardeau} \& {van de Weygaert}}{1996}]{Bernardeau:1996}
{Bernardeau} F.,  {van de Weygaert} R.,  1996, \mnras, 279, 693

\bibitem[\protect\citeauthoryear{{Bernardeau}, {van de Weygaert}, {Hivon} \&
  {Bouchet}}{{Bernardeau} et~al.}{1997}]{Bernardeau:1997}
{Bernardeau} F.,  {van de Weygaert} R.,  {Hivon} E.,    {Bouchet} F.~R.,  1997,
  \mnras, 290, 566

\bibitem[\protect\citeauthoryear{{Bertschinger} \& {Dekel}}{{Bertschinger} \&
  {Dekel}}{1989}]{Bertschinger:1989}
{Bertschinger} E.,  {Dekel} A.,  1989, \apjl, 336, L5

\bibitem[\protect\citeauthoryear{{Bode}, {Ostriker} \& {Turok}}{{Bode}
  et~al.}{2001}]{Bode:2001}
{Bode} P.,  {Ostriker} J.~P.,    {Turok} N.,  2001, \apj, 556, 93

\bibitem[\protect\citeauthoryear{{Buchert}}{{Buchert}}{1992}]{Buchert:1992}
{Buchert} T.,  1992, \mnras, 254, 729

\bibitem[\protect\citeauthoryear{{Carrasco}, {Foreman}, {Green} \&
  {Senatore}}{{Carrasco} et~al.}{2014}]{Carrasco:2014}
{Carrasco} J.~J.~M.,  {Foreman} S.,  {Green} D.,    {Senatore} L.,  2014,
  \jcap, 7, 57

\bibitem[\protect\citeauthoryear{{Carrasco}, {Hertzberg} \&
  {Senatore}}{{Carrasco} et~al.}{2012}]{Carrasco:2012}
{Carrasco} J.~J.~M.,  {Hertzberg} M.~P.,    {Senatore} L.,  2012, Journal of
  High Energy Physics, 9, 82

\bibitem[\protect\citeauthoryear{{Cautun} \& {van de Weygaert}}{{Cautun} \&
  {van de Weygaert}}{2011}]{Cautun:2011}
{Cautun} M.~C.,  {van de Weygaert} R.,  2011, arXiv:1105.0370

\bibitem[\protect\citeauthoryear{{Ciecielg} \& {Chodorowski}}{{Ciecielg} \&
  {Chodorowski}}{2004}]{Ciecielg:2004}
{Ciecielg} P.,  {Chodorowski} M.~J.,  2004, \mnras, 349, 945

\bibitem[\protect\citeauthoryear{{Colombi}, {Chodorowski} \&
  {Teyssier}}{{Colombi} et~al.}{2007}]{Colombi:2007}
{Colombi} S.,  {Chodorowski} M.~J.,    {Teyssier} R.,  2007, \mnras, 375, 348

\bibitem[\protect\citeauthoryear{{Dekel}}{{Dekel}}{1994}]{Dekel:1994}
{Dekel} A.,  1994, \araa, 32, 371

\bibitem[\protect\citeauthoryear{{Efstathiou}, {Davis}, {White} \&
  {Frenk}}{{Efstathiou} et~al.}{1985}]{Efstathiou:1985}
{Efstathiou} G.,  {Davis} M.,  {White} S.~D.~M.,    {Frenk} C.~S.,  1985,
  \apjs, 57, 241

\bibitem[\protect\citeauthoryear{{Eisenstein} \& {Hu}}{{Eisenstein} \&
  {Hu}}{1999}]{Eisenstein:1999}
{Eisenstein} D.~J.,  {Hu} W.,  1999, \apj, 511, 5

\bibitem[\protect\citeauthoryear{{Hahn} \& {Abel}}{{Hahn} \&
  {Abel}}{2011}]{Hahn:2011}
{Hahn} O.,  {Abel} T.,  2011, \mnras, 415, 2101

\bibitem[\protect\citeauthoryear{{Hahn}, {Abel} \& {Kaehler}}{{Hahn}
  et~al.}{2013}]{Hahn:2012}
{Hahn} O.,  {Abel} T.,    {Kaehler} R.,  2013, \mnras, 434, 1171

\bibitem[\protect\citeauthoryear{{Hahn}, {Carollo}, {Porciani} \&
  {Dekel}}{{Hahn} et~al.}{2007}]{Hahn:2007}
{Hahn} O.,  {Carollo} C.~M.,  {Porciani} C.,    {Dekel} A.,  2007, \mnras, 381,
  41

\bibitem[\protect\citeauthoryear{{Hahn}, {Porciani}, {Carollo} \&
  {Dekel}}{{Hahn} et~al.}{2007}]{HahnPorciani:2007}
{Hahn} O.,  {Porciani} C.,  {Carollo} C.~M.,    {Dekel} A.,  2007, \mnras, 375,
  489

\bibitem[\protect\citeauthoryear{{Henon}}{{Henon}}{1982}]{Henon:1982}
{Henon} M.,  1982, \aap, 114, 211

\bibitem[\protect\citeauthoryear{{Hockney} \& {Eastwood}}{{Hockney} \&
  {Eastwood}}{1981}]{Hockney:1981}
{Hockney} R.~W.,  {Eastwood} J.~W.,  1981, {Computer Simulation Using
  Particles}.
McGraw-Hill

\bibitem[\protect\citeauthoryear{{Hoffman}, {Metuki}, {Yepes}, {Gottl{\"o}ber},
  {Forero-Romero}, {Libeskind} \& {Knebe}}{{Hoffman}
  et~al.}{2012}]{Hoffman:2012}
{Hoffman} Y.,  {Metuki} O.,  {Yepes} G.,  {Gottl{\"o}ber} S.,  {Forero-Romero}
  J.~E.,  {Libeskind} N.~I.,    {Knebe} A.,  2012, \mnras, 425, 2049

\bibitem[\protect\citeauthoryear{{Icke} \& {van de Weygaert}}{{Icke} \& {van de
  Weygaert}}{1987}]{Icke:1987}
{Icke} V.,  {van de Weygaert} R.,  1987, \aap, 184, 16

\bibitem[\protect\citeauthoryear{{Jennings}}{{Jennings}}{2012}]{Jennings:2012}
{Jennings} E.,  2012, \mnras, 427, L25

\bibitem[\protect\citeauthoryear{{Jennings}, {Baugh} \& {Pascoli}}{{Jennings}
  et~al.}{2011}]{Jennings:2011}
{Jennings} E.,  {Baugh} C.~M.,    {Pascoli} S.,  2011, \mnras, 410, 2081

\bibitem[\protect\citeauthoryear{{Kitaura}, {Angulo}, {Hoffman} \&
  {Gottl{\"o}ber}}{{Kitaura} et~al.}{2012}]{Kitaura:2012}
{Kitaura} F.-S.,  {Angulo} R.~E.,  {Hoffman} Y.,    {Gottl{\"o}ber} S.,  2012,
  \mnras, 425, 2422

\bibitem[\protect\citeauthoryear{{Komatsu}, {Smith} \& {Dunkley}}{{Komatsu}
  et~al.}{2011}]{Komatsu:2011}
{Komatsu} E.,  {Smith} K.~M.,    {Dunkley} e.~a.,  2011, \apjs, 192, 18

\bibitem[\protect\citeauthoryear{{Kudlicki}, {Chodorowski}, {Plewa} \&
  {R{\'o}{\.z}yczka}}{{Kudlicki} et~al.}{2000}]{Kudlicki:2000}
{Kudlicki} A.,  {Chodorowski} M.,  {Plewa} T.,    {R{\'o}{\.z}yczka} M.,  2000,
  \mnras, 316, 464

\bibitem[\protect\citeauthoryear{{Laigle}, {Pichon}, {Codis}, {Dubois}, {Le
  Borgne}, {Pogosyan}, {Devriendt}, {Peirani}, {Prunet}, {Rouberol}, {Slyz} \&
  {Sousbie}}{{Laigle} et~al.}{2015}]{Laigle:2015}
{Laigle} C.,  {Pichon} C.,  {Codis} S.,  {Dubois} Y.,  {Le Borgne} D.,
  {Pogosyan} D.,  {Devriendt} J.,  {Peirani} S.,  {Prunet} S.,  {Rouberol} S.,
  {Slyz} A.,    {Sousbie} T.,  2015, \mnras, 446, 2744

\bibitem[\protect\citeauthoryear{{Libeskind}, {Hoffman}, {Steinmetz},
  {Gottl{\"o}ber}, {Knebe} \& {Hess}}{{Libeskind}
  et~al.}{2013}]{Libeskind:2013}
{Libeskind} N.~I.,  {Hoffman} Y.,  {Steinmetz} M.,  {Gottl{\"o}ber} S.,
  {Knebe} A.,    {Hess} S.,  2013, \apjl, 766, L15

\bibitem[\protect\citeauthoryear{{Melott}}{{Melott}}{1982}]{Melott:1982}
{Melott} A.~L.,  1982, Physical Review Letters, 48, 894

\bibitem[\protect\citeauthoryear{{Melott} \& {Shandarin}}{{Melott} \&
  {Shandarin}}{1993}]{Melott:1993}
{Melott} A.~L.,  {Shandarin} S.~F.,  1993, \apj, 410, 469

\bibitem[\protect\citeauthoryear{{Monaghan}}{{Monaghan}}{1992}]{Monaghan:1992}
{Monaghan} J.~J.,  1992, \araa, 30, 543

\bibitem[\protect\citeauthoryear{{Pandey}, {White}, {Springel} \&
  {Angulo}}{{Pandey} et~al.}{2013}]{Pandey:2012}
{Pandey} B.,  {White} S.~D.~M.,  {Springel} V.,    {Angulo} R.~E.,  2013,
  \mnras, 435, 2968

\bibitem[\protect\citeauthoryear{{Peebles}}{{Peebles}}{1980}]{Peebles:1980}
{Peebles} P.~J.~E.,  1980, {The large-scale structure of the universe}.
Princeton University Press

\bibitem[\protect\citeauthoryear{{Peebles}, {Melott}, {Holmes} \&
  {Jiang}}{{Peebles} et~al.}{1989}]{Peebles:1989}
{Peebles} P.~J.~E.,  {Melott} A.~L.,  {Holmes} M.~R.,    {Jiang} L.~R.,  1989,
  \apj, 345, 108

\bibitem[\protect\citeauthoryear{{Pelupessy}, {Schaap} \& {van de
  Weygaert}}{{Pelupessy} et~al.}{2003}]{Pelupessy:2003}
{Pelupessy} F.~I.,  {Schaap} W.~E.,    {van de Weygaert} R.,  2003, \aap, 403,
  389

\bibitem[\protect\citeauthoryear{{Pichon} \& {Bernardeau}}{{Pichon} \&
  {Bernardeau}}{1999}]{Pichon:1999}
{Pichon} C.,  {Bernardeau} F.,  1999, \aap, 343, 663

\bibitem[\protect\citeauthoryear{{Pueblas} \& {Scoccimarro}}{{Pueblas} \&
  {Scoccimarro}}{2009}]{Pueblas:2009}
{Pueblas} S.,  {Scoccimarro} R.,  2009, \prd, 80, 043504

\bibitem[\protect\citeauthoryear{{Pumir}, {Bodenschatz} \& {Xu}}{{Pumir}
  et~al.}{2013}]{Pumir:2013}
{Pumir} A.,  {Bodenschatz} E.,    {Xu} H.,  2013, Physics of Fluids, 25, 035101

\bibitem[\protect\citeauthoryear{{Schaap}}{{Schaap}}{2007}]{Schaap:2007}
{Schaap} W.~E.,  2007, PhD thesis, Kapteyn Astronomical Institute

\bibitem[\protect\citeauthoryear{{Schaap} \& {van de Weygaert}}{{Schaap} \&
  {van de Weygaert}}{2000}]{Schaap:2000}
{Schaap} W.~E.,  {van de Weygaert} R.,  2000, \aap, 363, L29

\bibitem[\protect\citeauthoryear{{Shandarin}, {Habib} \&
  {Heitmann}}{{Shandarin} et~al.}{2012}]{Shandarin:2011}
{Shandarin} S.,  {Habib} S.,    {Heitmann} K.,  2012, \prd, 85, 083005

\bibitem[\protect\citeauthoryear{{Sharma} \& {Steinmetz}}{{Sharma} \&
  {Steinmetz}}{2006}]{2006MNRAS.373.1293S}
{Sharma} S.,  {Steinmetz} M.,  2006, \mnras, 373, 1293

\bibitem[\protect\citeauthoryear{{Springel}}{{Springel}}{2005}]{Springel:2005}
{Springel} V.,  2005, \mnras, 364, 1105

\bibitem[\protect\citeauthoryear{{van de Weygaert} \& {Bond}}{{van de Weygaert}
  \& {Bond}}{2008}]{Weygaert:2008}
{van de Weygaert} R.,  {Bond} J.~R.,  2008, in {Plionis} M.,  {L{\'o}pez-Cruz}
  O.,   {Hughes} D.,  eds, A Pan-Chromatic View of Clusters of Galaxies and the
  Large-Scale Structure Vol.~740 of Lecture Notes in Physics, Berlin Springer
  Verlag, {Observations and Morphology of the Cosmic Web}.
p.~409

\bibitem[\protect\citeauthoryear{{Wang}, {Szalay}, {Arag{\'o}n-Calvo},
  {Neyrinck} \& {Eyink}}{{Wang} et~al.}{2014}]{Wang:2014}
{Wang} X.,  {Szalay} A.,  {Arag{\'o}n-Calvo} M.~A.,  {Neyrinck} M.~C.,
  {Eyink} G.~L.,  2014, \apj, 793, 58

\end{thebibliography}
